\documentclass[%
 reprint,
superscriptaddress,
longbibliography,
 amsmath,amssymb,
aps,
 prb,
floatfix,
]{revtex4-2}

\usepackage[T1]{fontenc}
\usepackage[utf8]{inputenc}
\usepackage[english]{babel}
\usepackage{stackengine}
\usepackage[]{graphicx}
\usepackage[caption=false]{subfig}
\usepackage{amsmath}
\usepackage{amssymb}
\usepackage{amsfonts}
\usepackage[]{units}
\usepackage{times}
\usepackage{physics}
\usepackage{bm}
\usepackage{bbold}
\usepackage{dsfont}
\usepackage{dutchcal}
\usepackage{xcolor}

\usepackage{tabularx}

\usepackage{hyperref}
\hypersetup{colorlinks=true, citecolor=blue, urlcolor=teal, linkcolor=blue}

\newcommand{\re}{\Re\textnormal{e}}
\newcommand{\im}{\Im\textnormal{m}}
\newcommand{\comment}[1]{}

\newcommand{\bQp}{\ensuremath{\mathbf{q}_\parallel}}
\newcommand{\Qp}{\ensuremath{q_\parallel}}
\newcommand{\bGp}{\ensuremath{\mathbf{g}_\parallel}}

\DeclareUnicodeCharacter{2212}{$-$} 
\DeclareUnicodeCharacter{2082}{$_2$} 

\usepackage{letltxmacro}
\LetLtxMacro{\ORIGselectlanguage}{\selectlanguage}
\makeatletter
\DeclareRobustCommand{\selectlanguage}[1]{%
  \@ifundefined{alias@\string#1}
    {\ORIGselectlanguage{#1}}
    {\begingroup\edef\x{\endgroup
       \noexpand\ORIGselectlanguage{\@nameuse{alias@#1}}}\x}%
}
\newcommand{\definelanguagealias}[2]{%
  \@namedef{alias@#1}{#2}%
}

\makeatother

\definelanguagealias{en}{english}
\definelanguagealias{EN}{english}
\definelanguagealias{eng}{english}
\definelanguagealias{de}{ngerman}

\begin{document}

\title{Strong Coupling of Two-Dimensional Excitons and Plasmonic Photonic Crystals:\\ Microscopic Theory Reveals Triplet Spectra}

\author{Lara Greten}
\email{lara.greten@tu-berlin.de}
\affiliation{Nichtlineare Optik und Quantenelektronik, Institut f\"ur Theoretische Physik, Technische Universit\"at Berlin,  10623 Berlin, Germany}
\author{Robert Salzwedel}
\affiliation{Nichtlineare Optik und Quantenelektronik, Institut f\"ur Theoretische Physik, Technische Universit\"at Berlin,  10623 Berlin, Germany}
\author{Tobias Göde}
\affiliation{Nichtlineare Optik und Quantenelektronik, Institut f\"ur Theoretische Physik, Technische Universit\"at Berlin,  10623 Berlin, Germany}
\author{David Greten}
\affiliation{Formerly: Institut f\"ur Theoretische Physik, Technische Universit\"at Berlin,  10623 Berlin, Germany \\Current address: Fritz Haber Institute of the Max Planck Society, Theory Department, 14195 Berlin, Germany}
\author{Stephanie Reich}
\affiliation{Experimentelle Festkörperphysik, Freie Universität Berlin, 14195 Berlin, Germany}
\author{Stephen Hughes}
\affiliation{Department of Physics, Engineering Physics and Astronomy, Queen’s University, Kingston, Ontario K7L 3N6, Canada}
\author{Malte Selig}
\affiliation{Nichtlineare Optik und Quantenelektronik, Institut f\"ur Theoretische Physik, Technische Universit\"at Berlin,  10623 Berlin, Germany}
\author{Andreas Knorr}
\affiliation{Nichtlineare Optik und Quantenelektronik, Institut f\"ur Theoretische Physik, Technische Universit\"at Berlin,  10623 Berlin, Germany}

\begin{abstract}
Monolayers of transition metal dichalcogenides (TMDC) are direct-gap semiconductors with strong light-matter interactions featuring tightly bound excitons, while plasmonic crystals (PCs), consisting of metal nanoparticles that act as meta-atoms, exhibit collective plasmon modes and allow one to tailor electric fields on the nanoscale.
Recent experiments show that TMDC-PC hybrids can reach the strong-coupling limit between excitons and plasmons forming new quasiparticles, so-called plexcitons. To describe this coupling theoretically, we develop a self-consistent Maxwell-Bloch theory for TMDC-PC hybrid structures, which allows us to compute the scattered light in the near- and far-field explicitly and provide guidance for experimental studies.
Our calculations 
reveal a spectral splitting signature of strong coupling of more than~$100\,$meV in gold-MoSe$_2$ structures with $30\,$nm nanoparticles, manifesting in a hybridization of exciton and plasmon into two effective plexcitonic bands.
In addition to the hybridized states, we find a remaining excitonic mode with significantly smaller coupling to the plasmonic near-field, emitting directly into the far-field. Thus, hybrid spectra in the strong coupling regime can contain three emission peaks.
\end{abstract}

\maketitle


\section{Introduction}
The light-matter interaction strength in transition metal dichalcogenide (TMDC) monolayers has been reported to be extremely strong~\cite{kusch_strong_2021}, e.g., as demonstrated by absorption rates of up to 10\% in the visible spectrum~\cite{mak_atomically_2010,splendiani_emerging_2010}. Such a high absorption is particularly noteworthy given the two-dimensional nature of these materials, which possess a thickness of less than $1\,$nm. 
In addition to featuring a direct bandgap~\cite{mak_atomically_2010}, TMDC monolayers support in-plane exciton formation due to their two-dimensional structure~\cite{haug_quantum_2004}. Excitons (bound electron-hole pairs) therefore dominate the optical spectrum below the band edge~\cite{wang_colloquium_2018}.
In addition, the remarkably thin nature of TMDC monolayers results in an increased sensitivity to surrounding materials. Consequently, the atomically thin materials can easily be influenced by various factors, such as the choice of the substrate material, defects~\cite{greben_intrinsic_2020,mitterreiter_role_2021}, and functionalization \cite{hu_polariton_2021}, e.g., with molecules \cite{feierabend_proposal_2017,katzer_impact_2023} and heterostructure configurations \cite{selig_theory_2019,park_temperaturedependent_2021}.

\begin{figure}[b]
 \begin{center}
 \includegraphics[width=\linewidth]{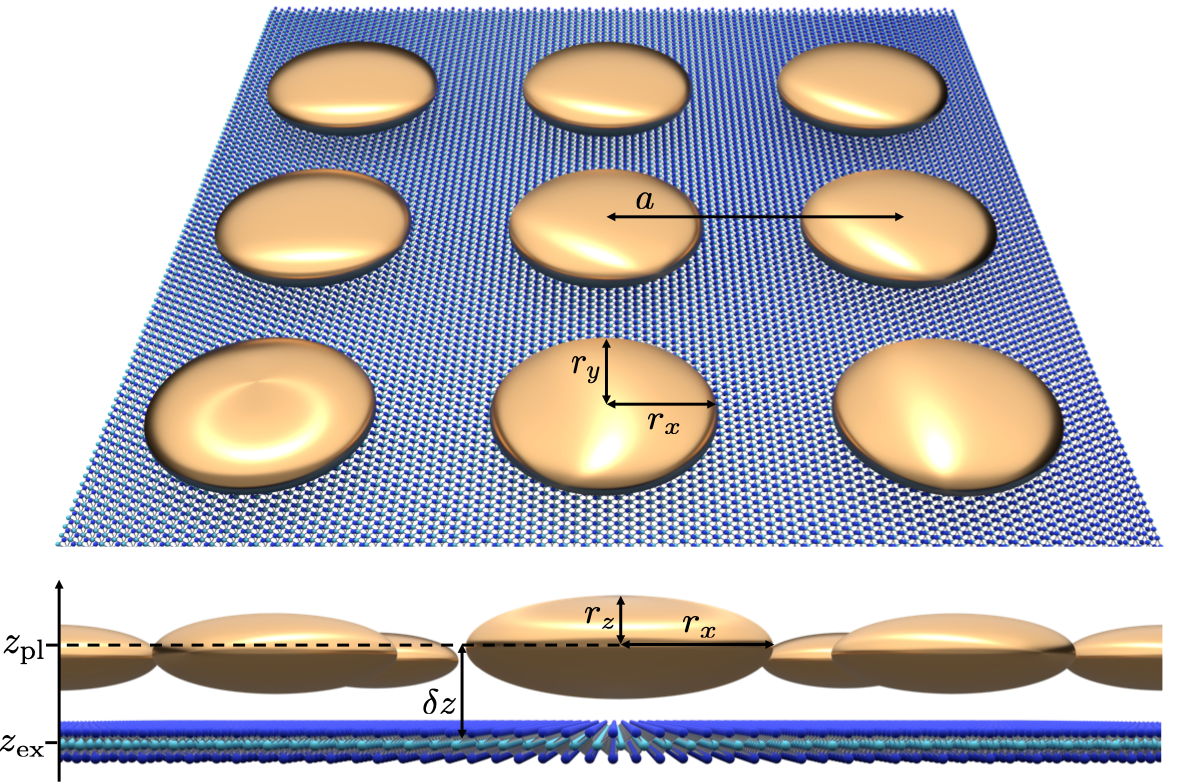}
 \end{center}
 \caption{\textbf{Sketch of the hybrid system:} the 2D semiconductor (TMDC) is covered by a square-structured 2D PC of metal nanodisks with the example of gold. The structure is periodic and infinite in the $xy$-plane and embedded in a surrounding medium with constant permittivity $\varepsilon$.}
 \label{fig: sketch hybrid}
\end{figure}

In contrast, the optical response of metal nanoparticles (MNPs) is dominated by localized plasmons which are collective electron oscillations formed within the metal conduction band~\cite{maier_plasmonics_2007}.
A special feature of MNPs is a significant amplification of the electric near-field, which additionally allows for manipulating the electric field on dimensions far below the diffraction 
limit~\cite{tame_quantum_2013,maier_plasmonicsroute_2001}. Arranging MNPs as meta-atoms in a crystal structure yields a plasmonic crystal (PC) with extraordinary strong light-matter interaction. The localized plasmons couple with the electric field and form plasmon-polaritons which can propagate within the crystal~\cite{lamowski_plasmon_2018, mueller_deep_2020} and sharpen the single particle plasmon resonance.
The optical properties of PCs strongly depend on a variety of parameters like lattice structure and nanoparticle shape. By manipulating these, it is possible to tune the optical properties of the crystal over a wide range \cite{lamowski_plasmon_2018,schulz_structural_2020,wang_manipulating_2019, epishin_theory_2023}. The strong tunability and enhancement of the electric field make periodic plasmonic structures appealing for light harvesting and non-linear optics, yielding, e.g., applications for nanoscale lasing~\cite{zhou_lasing_2013,wang_manipulating_2019} and advanced optical spectroscopy \cite{mueller_surface-enhanced_2021,langer_present_2020}.

It has also been shown that the absorption of graphene is significantly enhanced by depositing plasmonic nanostructures near the graphene layer~\cite{mueller_microscopic_2018,echtermeyer_strong_2011}.
An interaction between localized surface plasmons of a single MNP and excitons in the semiconductor plane can reach the strong coupling regime~\cite{salzwedel_spatial_2023,carlson_strong_2021, petric_tuning_2022, kleemann_strong-coupling_2017, geisler_single-crystalline_2019, wen_room-temperature_2017, zheng_manipulating_2017,goncalves_plasmon-exciton_2018} and absorption rates up to $90\%$ \cite{gomez_near-perfect_2021}.
This motivates that combining the particular properties of PCs with the environment-sensitive semiconductor monolayer promises a further increase of the light-matter-interaction of the excitons in the TMDC.

In the context of our study, we adopt a definition of ``strong coupling'' where the interaction strength between the two systems\cite{khitrova_nonlinear_1999,jahnke_excitonic_1996} exceeds the total losses of the system, 
where the former is quantified by an effective Rabi splitting $\Omega_\text{eff}\approx 2g_\text{eff}>\gamma_\text{ex}+\gamma_\text{pl}$~\cite{zheng_manipulating_2017,wen_room-temperature_2017,schneider_two-dimensional_2018} with $g_\text{eff}$ an effective exciton-plasmon interaction strength.
In a classical theory, this is usually termed normal mode splitting, though the underlying dissipative modes are quasinormal modes with complex eigenfrequencies, which are well characterized by Maxwell's equations~\cite{carlson_strong_2021}.
This definition is particularly relevant for exciton-plasmon coupling as the losses in the plasmonic components are significantly larger than the losses in the excitonic component~\cite{carlson_strong_2021}.
In this work, we study the impact of a two-dimensional (2D) PC on the excitonic dynamics in TMDCs.
A sketch of the hybrid system is depicted in Fig.~\ref{fig: sketch hybrid}. The semiconductor is located parallel to the $xy$-plane. The 2D PC of gold disks (square array) is placed on top of the TMDC. 
Experimentally, a similar configuration for relatively large nanoparticles with an in-plane radius of more than $50$ nm was already realized in Refs.~\cite{abid_temperature-dependent_2017,liu_strong_2016,vadia_magneto-optical_2023,zhang_observation_2023}, where TMDCs are coupled to 2D PCs with different lattice structures.

To theoretically study the exciton-plasmon coupling, we develop a self-consistent Maxwell-Bloch theory for the hybrid structure in Sec.~\ref{SecTheo}.
We first give a short review on the solution of the Maxwell's equations (Sec.~\ref{sec: Maxwells equations}), the description of the TMDC monolayer with the excitonic Bloch equation (Sec.~\ref{sec: excitons}) and the 2D PC using Mie theory (Sec.~\ref{sec: Plasmon}).
This allows us in Sec.~\ref{sec: plexciton} to use the plasmonic dipole density within the PC as a source for the excitonic dipole density of the TMDC, by inserting it into the excitonic Bloch equations. The resulting formulas are Bloch equations that describe the dynamics of excitons in the presence of an electric field and the field-mediated interaction. We highlight the qualitatively different behavior of momentum-dark excitons, that couple to the plasmon-enhanced electric near-field, and originally bright excitons with vanishing center-of-mass momentum.
Finally, an expression for the emitted electric field is derived from which we can deduce transmission, reflection, and absorption of the hybrid structure when using a plane wave excitation.

Using this theoretical framework, we perform numerical calculations to obtain the absorption spectra of the hybrid system in Sec.~\ref{sec: results}. Our analysis resembles the broad range of coupling regimes, varying from weak to strong.
In the strong coupling case, we find an additional peak at the unperturbed exciton energy, which exhibits a strong dependence with the temperature and coupling strength. Our theory suggests that this additional mode stems from the originally bright excitons that do not participate in the strong coupling to the plasmon-enhanced electric near-field. We find that this excitonic mode is uniformly distributed in real space, thus also located in the immediate vicinity of the MNP. The observed strong coupling between TMDCs and MNPs or PCs is based on the coupling between plasmon and momentum-dark excitons.
We note that there is ongoing discussion regarding exciton-plasmon coupling in nanoshells, with some studies also suggesting the existence of an undisturbed excitonic mode~\cite{antosiewicz_plasmon-exciton_2014}. To the best of our knowledge, this mode has not been observed in experiments for nanoshells~\cite{stete_optical_2023}.
However, our results agree with recent experiments~\cite{vadia_magneto-optical_2023} which observe the presence of such an additional excitonic peak for TMDC-PC hybrids. 

\section{Theoretical Model}\label{SecTheo}
In this section, we develop a theoretical framework that describes the interaction between TMDC excitons and PC plasmons via the radiation field.
\subsection{Maxwell's Equations\label{sec: Maxwells equations}}
Starting from Maxwell's equations, assuming a non-magnetic and isotropic medium, 
the wave equation for the electric field reads
\begin{align}
\left(\boldsymbol{\nabla}^2-\frac{\varepsilon}{c^2}{\partial_t^2}\right)\mathbf{E}(\mathbf{r},t)&=
 \mathbf{S}(\mathbf{r},t)\label{eq: wave equation},
\end{align}
with the source
 \begin{align}
 \mathbf{S}(\mathbf{r},t) &= \left(\frac{\partial_t^2}{\varepsilon_0c^2}
-\frac{1}{\varepsilon_0\varepsilon}\boldsymbol{\nabla} \boldsymbol{\nabla}\cdot\right) \mathbf{P}(\mathbf{r},t)\label{eq: wave equation source},
\end{align}
which is valid for a freestanding sample with dipole density $\mathbf{P}(\mathbf{r},t)$ embedded in a homogeneous, isotropic and non-dispersive dielectric environment with permittivity $\varepsilon$.
A solution of Eq.~\eqref{eq: wave equation} can be formally obtained via the scalar Green's function $G(\mathbf{r}, \mathbf{r^\prime} , t-t^\prime)$,
\begin{align}
    \mathbf{E}(\mathbf{r},t) =\int dt^\prime
\int d^3r^\prime\, G(\mathbf{r}, \mathbf{r^\prime} ,t-t^\prime)\,
\mathbf{S}(\mathbf{r}^\prime,t^\prime),\label{eq: wave equation, formal sol. 1}
\end{align}
where the scalar Green's function depends on the boundary conditions.
Both layers of the hybrid structure, cf.~Fig.~\ref{fig: sketch hybrid}, are assumed to be aligned parallel to the $xy$-plane with a discrete translation invariance for the PC. We treat the $z$-direction separately and apply a Fourier transform in the in-plane coordinates $x$, $y$ and time $t$,
\begin{align}
    \mathbf{E}_{\bQp}(z,\omega) =
\int dz^\prime\,
\mathbcal{G}_{\bQp}(z,z^\prime,\omega)\cdot
\mathbf{P}_{\bQp}(z^\prime,\omega),\label{eq: wave equation, formal sol. 2}
\end{align}
where we introduced the new Green's dyadic $\mathbcal{G}_{\bQp}(z,z',\omega)$ that converts the dipole density at the position $z^\prime$ into an electric field and describes its propagation from $z^\prime$ to the observation position $z$.
The term
$\mathbcal{G}_{\bQp}$ can be derived analytically~\cite{sipe_new_1987,tomas_green_1995},
\begin{align}
&\mathbcal{G}_{\bQp}(z,z',\omega) = \nonumber \\
 & \ \ \begin{pmatrix}
-\frac{\omega^2}{\varepsilon_0c^2}\mathbb{1} + \frac{\bQp\otimes\,\bQp}{\varepsilon_0\varepsilon}
 & \frac{i\bQp} {\varepsilon_0\varepsilon}\partial_{z'} \\ & \\
\frac{i\bQp^T}{\varepsilon_0\varepsilon}\partial_{z'} &
-\frac{\omega^2}{\varepsilon_0c^2} - \frac{1}{\varepsilon_0\varepsilon}\partial_{z'}^2 
\end{pmatrix}
G_{\Qp} (z,z',\omega). \label{eq:dyadic Green's function}
\end{align}
Equation \eqref{eq:dyadic Green's function} gives the Green's dyadic in Cartesian coordinates regarding the $z$-direction whereas the in-plane components are expressed independent of a special basis for $\bQp$, which will be specified in Sec.~\ref{sec: plexciton}.
The scalar Green's function for a constant surrounding permittivity is
\begin{align}
G_{\Qp}(z,z',\omega)&=\frac{-i}{2k}e^{ik_{\bQp}|z-z'|},
\hspace{0.2 cm}\hspace{0.2 cm}
k_{\bQp}=\sqrt{\frac{\varepsilon}{c^2}\omega^2-\Qp^2},\label{eq: Green's function, omega dependent}
\end{align}
where we abbreviate the absolute value of a vector by, e.g., $\vert \bQp \vert = \Qp$.

Next, using the thin film approximation for the individual layers~\cite{knorr_theory_1996,stroucken_coherent_1996}, i.e., with TMDC (ex) and PC (pl), then
\begin{equation}
\mathbf{P}_{\bQp}(z,\omega)=  
\sum_{\ell= \text{\{ex,pl\}}\hspace{-0.6cm}}
\mathbf{P}^{2\text{D},\ell}_{\bQp}(\omega)\,
\delta (z-z_\ell)\label{eq: thin film approximation}
\end{equation}
allows to find an algebraic solution of Maxwell's equations in Fourier space.
Adding the incident field $\mathbf{E}^0_{\bQp}(z,\omega)$ as a solution of the homogeneous wave equation, Eq.~\eqref{eq: wave equation}, yields 
\begin{equation}
\mathbf{E}_{\bQp}(z,\omega) = \sum_{\ell= \text{\{ex,pl\} }\hspace{-0.6cm} }\mathbcal{G}_{\bQp}(z,z_\ell,\omega)\cdot\mathbf{P}^{2\text{D},\ell}_{\bQp}(\omega)+\mathbf{E}^0_{\bQp}(z,\omega), \label{eq: full electric field start}
\end{equation}
which allows us to compute the 
transmission and reflection.
Thus, we can connect the dipole density to experimentally measurable far-field signals.

We assume a perpendicular plane wave excitation that is propagating in the positive $z$-direction,
\begin{align}
    \mathbf{E}^0_{\bQp}(\omega,z)=e^{ikz}\delta(\bQp)\,\mathbf{E}^0(\omega),\label{eq: plane wave E0}
\end{align}
with an amplitude perpendicular to the $z$-axis $\mathbf{E}^0\perp \mathbf{e}_z$.
In the far-field limit, only a $\Qp=0$ Fourier component occurs, see Eq.~\eqref{eq: dense PC approximation} . The transmission is given by
\begin{align}
  &T(\omega)
=\frac{\vert\mathbf{E}_{{\bQp}=\mathbf{0}}(z,\omega)\vert^2}{\vert\mathbf{E}^0_{{\bQp}=\mathbf{0}}(z,\omega)\vert^2 },
\hspace{0.2 cm} \,z\rightarrow \infty,
\label{eq: transmission}
\end{align}
and the reflection is
\begin{align}
  &R(\omega)
=\frac{\vert\mathbf{E}_{{\bQp}=\mathbf{0}}(z,\omega)-\mathbf{E}^0_{{\bQp}=\mathbf{0}}(z,\omega)\vert^2}{\vert\mathbf{E}^0_{{\bQp}=\mathbf{0}}(z,\omega)\vert^2 },\hspace{0.2 cm} \,z\rightarrow -\infty.
\label{eq: reflection}
\end{align}
The
absorption is then easily obtained from
\begin{align}
    A(\omega)=1-T(\omega)-R(\omega)\label{eq: TRA 1}.
\end{align}

\subsection{Excitonic Dipole Density\label{sec: excitons}}
To describe the response of the TMDC excitons to the electric field, we define the macroscopic 2D dipole density~\cite{katsch_exciton-scattering-induced_2020,knorr_theory_1996}
\begin{equation}
\mathbf{P}^{2\text{D,ex}}_{{\bQp}}(\omega)= \sum\limits_{\xi} \mathbf{d}^{\xi}\varphi_{\mathbf{r}_\parallel = \mathbf{0}} \, p^{\xi}_{{\bQp}}(\omega), \label{eq: definition, TMDC polarization}
\end{equation}
with the dipole moment $\mathbf{d}^\xi$, $\xi = +/-$ corresponding to the $K^+$ and $K^-$ valley, carrying the circular dichroism~\cite{xiao_coupled_2012}. The strength of the dipole moment $\mathbf{d}^\xi$ is taken from DFT calculations~\cite{xiao_coupled_2012}; the term $\varphi_{\mathbf{r}_\parallel = \mathbf{0}}$ accounts for the value of the 1s excitonic wave function in real space at $\mathbf{r}_\parallel = \mathbf{0}$ and is obtained from the solution of the Wannier equation
\cite{berghauser_analytical_2014,selig_excitonic_2016}, incorporating the dielectric environment in the Rytova-Keldysh approximation~\cite{rytova_screened_1967}.
We concentrate on the lowest 1s excitonic state, since it is energetically separated from higher transitions~\cite{wang_colloquium_2018}.
A table with all parameters for MoSe$_2$ can be found in Appendix A.
The excitonic transition amplitude is denoted as $p^{\xi}_{{\bQp}}$ and obtained via its Bloch equation in the Fourier domain~\cite{kormanyos_k_2015},
\begin{equation}
\left( \hbar\omega - \mathcal{E}_{\bQp} + i\gamma(T) \right) \mathbf{p}_{{\bQp}}(\omega)= 
-  {\mathbcal{d}}^* \cdot \mathbf{E}_{{\bQp}}(z_\text{ex},\omega)\label{eq:excitonicBlochEquation},
\end{equation}
with the dipole tensor
\begin{align}
  \mathbcal{d}=\varphi_{\mathbf{r}_\parallel = \mathbf{0}}\left( \mathbf{d}^{+},\mathbf{d}^{-},\mathbf{0}\right),
    \label{eq: definition dipole tensor}
\end{align}
and the excitonic transition in vector notation,
\begin{align}
    \mathbf{p}_{{\bQp}}(\omega)=\left( 
    \begin{array}{c}
        p^{+}_{{\bQp}} (\omega)   \\
        p^{-}_{{\bQp}} (\omega) \\
        0
    \end{array}
    \right).
\end{align}
The zero in the $z$ component is added for convenience, to only allow for the appearance of square matrices.

A detailed derivation is provided in Ref.~\cite{katsch_theory_2018}, where the rotating wave 
approximation (RWA) is utilized. This approximation imposes a constraint on the light-matter interaction strength, which must be significantly smaller than the system energies. Therefore, the range of exciton-plasmon coupling strength that can be accurately described in this work is limited to $g_\text{eff}/E^\text{pl}<0.1$ which is below the ultra-strong coupling regime~\cite{forn-diaz_ultrastrong_2019,frisk_kockum_ultrastrong_2019,mueller_deep_2020}.

The temperature-dependent dephasing rate $\gamma(T)$ accounts for non-radiative decay, which typically
results from exciton-phonon interactions and is calculated microscopically~\cite{selig_excitonic_2016},
\begin{align}
    \gamma(T) = c_1T+\frac{c_2}{\frac{\Omega}{e^{k_BT}}-1},
\end{align}
where $c_1$, $c_2,$ and the averaged phonon-energy $\Omega$ are given in Appendix A.
All radiative corrections such as lineshift/splitting and broadening are incorporated via 
the self-consistently calculated electric field $\mathbf{E}_{{\bQp}} (z_\text{ex},\omega)$ given in Eq.~\eqref{eq: full electric field start}.

\begin{figure}[t]
 \begin{center}
 \includegraphics[width=.9\linewidth, trim={0 2cm 0 0}]{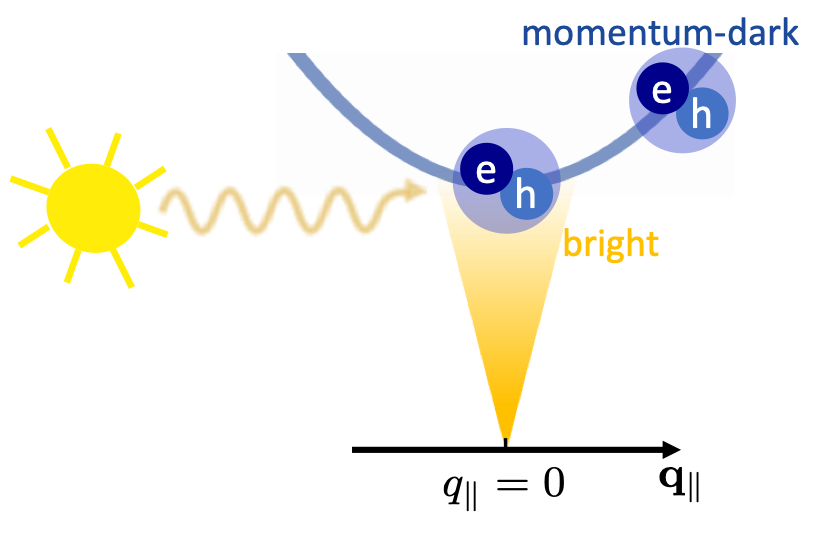}
 \end{center}
 \caption{\textbf{Parabolic exciton dispersion} over center-of-mass momentum ${\bQp}$. A bright and a momentum-dark exciton are indicated exemplary. The schematic ``sun'' represents a radiative excitation from the far-field.}
 \label{fig: exciton dispersion}
\end{figure}

The left-hand side of Eq.~\eqref{eq:excitonicBlochEquation} accounts for the dispersion of excitons
\begin{align}
\mathcal{E}_{\bQp} = E^\text{ex}+\frac{\hbar^2\bQp^2}{2M},\label{eq: exciton dispersion}
\end{align}
with center-of-mass wave number ${\bQp}$ (also referred to as in-plane momentum), the excitonic mass $M$, the excitonic transition energy $E^\text{ex}$ of 1s-excitons with vanishing momentum.
The parabolic exciton dispersion is illustrated in Fig.~\ref{fig: exciton dispersion}. In Eq.~\eqref{eq:excitonicBlochEquation}, an excitonic transition $\mathbf{p}_{{\bQp}}$ can be excited by an electric field $\mathbf{E}_{{\bQp}}$ with the same in-plane momentum. The excitonic transition $\mathbf{p}_{{\bQp}}$ is called \textit{bright} if it can be excited from the far-field. However, propagating solutions of the wave equation for the electric field, Eq.~\eqref{eq: wave equation}, are only possible for real values of $k_{\bQp}$, cf.~Eq.~\eqref{eq: Green's function, omega dependent}, meaning for small momenta $\Qp$. We refer to excitons with higher momenta as \textit{momentum-dark}, as they are inaccessible via far-field illumination.

To connect the solution of Eq.~\eqref{eq:excitonicBlochEquation} with the Green's dyadic formalism solving Maxwell's equations in the circular basis $(+,-)$, we adjust the notation of Eq.~\eqref{eq: definition, TMDC polarization} to 
\begin{equation}
\mathbf{P}^{2\text{D,ex}}_{{\bQp}}(\omega)= \mathbcal{d}\cdot \mathbf{p}_{{\bQp}}(\omega). \label{eq: matrix definition, TMDC polarization}
\end{equation}
Due to the circular dichroism of TMDCs, the dipole tensor is diagonal in a circular polarized basis,  $\mathbcal{d}^{\sigma\xi}=d\delta_{\sigma\xi}$.

\subsection{Plasmonic Polarizability\label{sec: Plasmon}}
To describe the response of MNPs, we use Mie-Gans theory~\cite{mie_beitrage_1908,gans_uber_1912}, providing a parametric frequency dependent polarizability $\boldsymbol{\alpha} (\omega)$. The optical response of every MNP is approximated by a point dipole $\mathbf{p}^j$ at the lattice position $\mathbf{R}^j$,
\begin{equation}
\mathbf{P}^{2\text{D,pl}}(\mathbf{r}_\parallel,\omega)=\sum_j \mathbf{p}^j(\omega)\,\delta(\mathbf{r}_\parallel-\mathbf{R}^j_\parallel)
\label{eq: plasmon dipole density},
\end{equation}
with
\begin{equation}
\mathbf{p}^j(\omega) = \boldsymbol{\alpha}(\omega)\cdot \mathbf{E}(\mathbf{R}^j,\omega) ,\label{eq: single nano-particle dipole}
\end{equation}
where $\mathbf{E}(\mathbf{R}^j,\omega)$ is the electric field, {\it excluding} the field generated by the MNP itself which is already incorporated in $\boldsymbol{\alpha}(\omega)$. The quasi-static polarizability $\boldsymbol{\alpha}^\text{qs}$ for spheroids is given by, e.g.,  
Mie-Gans theory~\cite{gans_uber_1912,bohren_absorption_1983}. It can be used for a quasi-static description, if the extensions of the MNPs are significantly smaller than the wavelength of the exciting light-field $r_i\ll \lambda$. In Cartesian coordinates, with the MNP axes along the corresponding semi-axes $(i=\{x,y,z\})$, the quasi-static polarizability is diagonal with the entries
\begin{equation}
    \alpha^\text{qs}_{ii}=4\pi\varepsilon_0\varepsilon\, r_xr_yr_z \frac{\varepsilon_\text{Au}(\omega)-\varepsilon}{3\varepsilon+3L_i(\varepsilon_\text{Au}(\omega)-\varepsilon)}, \label{eq: static polarizability single spheroid}
\end{equation}
where $r_i$ is the half-axis of the MNP along the corresponding direction, $\varepsilon_\text{Au}(\omega)$ is the permittivity of gold, $\varepsilon$ the surrounding permittivity and $L_i$ the shape factor for oblate spheroids,
\begin{align}
      L_x&=L_y= \frac{1}{2e_0}
    \left[ 
    \frac{\sqrt{1-e_0^2}}{e_0}\arcsin (e_0)- (1-e_0^2)
    \right] , \\
    L_z&= 1-2L_x,
\end{align}
with the eccentricity,
\begin{equation}
    e_0 = 1-\frac{r_z^2}{r_x^2}, \hspace{0.2cm}\text{if}\hspace{0.2cm} r_x=r_y.
\end{equation}

To construct the optical response of the MNP-based 2D PC depicted in Fig.~\ref{fig: sketch hybrid}, we need to take the radiative coupling between the nanoparticles into account. Therefore, we consider corrections to the quasi-static Mie-Gans solution of the single particle polarizability in the \textit{modified long-wavelength approximation} (MLWA) including correction terms for dynamic depolarization and radiation damping~\cite{van_vlack_spontaneous_2012,humphrey_plasmonic_2014,moroz_depolarization_2009},
\begin{equation}
    \alpha^\text{MLWA}_{ii}=\frac{\alpha^\text{qs}_{ii}}{1-\frac{2}{3} i\, k_{{\bQp}=0}^3\frac{\alpha^\text{qs}_{ii}}{4\pi\varepsilon_0\varepsilon}-\frac{k_{{\bQp}=0}^2}{r_i}\frac{\alpha^\text{qs}_{ii}}{4\pi\varepsilon_0\varepsilon}},
\end{equation}
with the radiative wavevector $k_{{\bQp}=0}=\sqrt{\epsilon}\frac{\omega}{c}$, compare for Eq.~\eqref{eq: Green's function, omega dependent}.
A full self-consistent solution would inherently include the radiation damping. However, in Eq.~\eqref{eq: single nano-particle dipole} the electric field generated by the MNP itself is excluded, making this correction necessary.
Whether dynamic depolarization and radiation damping are important for a single MNP depends on the size of the MNP: for particle diameters significantly smaller than $\lambda / 2\pi$, the static Mie-Gans theory, Eq.~\eqref{eq: static polarizability single spheroid}, would be a sufficient description. In Sec.~\ref{sec: results}, we evaluate the theory for MNPs with in-plane radii of $30\,$nm, where corrections due to the finite extent become essential for the description of their optical properties.

For the gold-MNP permittivity $\varepsilon_\text{Au}(\omega)$, as occurring in Eq.~\eqref{eq: static polarizability single spheroid}, a heuristic, analytical model that fits experimental data for bulk ~\cite{johnson_optical_1972} is parameterized by~\cite{etchegoin_analytic_2006},
\begin{align}
\varepsilon_\text{Au}(\omega)  =&\,\varepsilon_\infty-\frac{\omega^2_\text{p}}{\omega^2+i\Gamma(\omega,T)\omega}\nonumber\\
&+\sum_{j=1,2} A_j\omega_j\left( \frac{e^{i\phi_j}}{\omega_j-\omega-i\Gamma_j}+\frac{e^{-i\phi_j}}{\omega_j+\omega+i\Gamma_j}  \right). \label{eq: Au permittivity}
\end{align}
In the high-frequency limit, the permittivity $\varepsilon_\infty$ of gold differs from unity since $d$-bands in noble metals are filled and provide a high residual polarization \cite{maier_plasmonics_2007}.
Equation~\eqref{eq: Au permittivity} incorporates a Drude-like intraband response of the conduction band and two interband transitions with the possibility to fit asymmetric line shapes in linear bulk spectra (last two terms in Eq.~\eqref{eq: Au permittivity}). We combine the room-temperature permittivity given in Ref.~\cite{etchegoin_analytic_2006} with a temperature and spectrally dependent linewidth \cite{liu_reduced_2009,mckay_temperature_1976,parkins_intraband_1981,bouillard_low-temperature_2012} via the damping term $\Gamma(\omega,T)$ as a sum of electron-electron and electron-phonon scattering using a Debye model,
\begin{align}
    \Gamma (\omega,T) = \Gamma_{\text{el-el}}(\omega,T) + \Gamma_{\text{el-ph}}(T),\label{eq: drude damping}
\end{align}
with~\cite{liu_reduced_2009,mckay_temperature_1976,parkins_intraband_1981,bouillard_low-temperature_2012}
\begin{align}
    \Gamma_{\text{el-el}}(\omega,T) &= b \left[ (k_BT)^2 + (\hbar \omega/2\pi)^2\right]\label{eq: drude damping - elel},\\
    \Gamma_{\text{el-ph}}(T) &= \gamma_0\left[ \frac{2}{5}+ 4\left(\frac{T}{\Theta}\right)^5\int_0^{\Theta/T} \frac{z^4}{e^z-1}dz\right].\label{eq: drude damping - elph}
\end{align}
A table with all parameters, including the Debye temperature $\Theta$, for the permittivity of gold is given in Appendix A.
We consider the temperature dependence of the linewidth only for the Drude contribution in Eq.~\eqref{eq: Au permittivity}, since it is the main contribution in the well-studied energy regime below the energy range where interband transitions become relevant, $\hbar\omega< 2.4\,$eV \cite{bouillard_low-temperature_2012}.
For simplicity, we disregard a subtle redshift of $\omega_\text{p}$ for low temperatures.
Equation~\eqref{eq: drude damping} does not consider radiative damping, since it is included by self-consistently solving Maxwell's equations.
Equations~\eqref{eq: single nano-particle dipole}-\eqref{eq: Au permittivity} fully describe the response of single nanoparticles to a self-consistently calculated electric field $\mathbf{E}$, Eq.~\eqref{eq: full electric field start}.

In contrast to the single MNP properties, in a PC, the field generated by all other nanoparticles has to be considered self-consistently.
The discrete translational invariance in the $xy$-plane allows to collect the interactions with the other MNPs in a modified polarizability $\boldsymbol{\alpha}^*_{\bQp}(\omega)$ in Fourier space. A self-consistent solution for the two-dimensional dipole density of the 2D PC in coupled dipole approximation is given by Ref.~\cite{garcia_de_abajo_colloquium_2007} and experimentally confirmed for gold nanodisks with $r_x=r_y\approx60\,$nm and a lattice constant $a\approx 500\,$nm in Ref.~\cite{humphrey_plasmonic_2014}. The corresponding dipole density reads
\begin{equation}
\mathbf{P}^{2\text{D,pl}}_{{\bQp}}(\omega)=\boldsymbol{\alpha}^*_{\bQp}(\omega)\cdot\sum\limits_{\bGp} \mathbf{E}_{\bQp+{\bGp}}(z_\text{pl},\omega)\label{eq: plasmon polarization},
\end{equation}
where the sum includes all reciprocal lattice vectors ${\bGp}$ corresponding to Umklapp processes.
The effective polarizability tensor $\boldsymbol{\alpha}^*_{\bQp}$ contains corrections due to interactions between the nanoparticles in the coupled dipole approximation restricted to MNP center-to-center distances $a\geq 3r_{xy}$. In this limit~\cite{brongersma_electromagnetic_2000}, the effective polarizability is given by
\begin{equation} 
\boldsymbol{\alpha}^*_{\bQp}(\omega)
 =\frac{1}{A^\text{pl}_\text{UC}}
 \left( \left(\boldsymbol{\alpha}^\text{MLWA}(\omega)\right)^{-1} - \frac{1}{4\pi\epsilon\epsilon_0}\mathbcal{F}_{\bQp}(\omega)\right)^{-1}\label{eq:interaction corrected polarizability},
\end{equation}
normalized by the unit cell area $A^\text{pl}_\text{UC}$ of the PC.
The form factor
\begin{align}
    &\mathbcal{F}_{\bQp}(\omega)= \label{eq: gold scattering matrix S}\\
    &
    \sum_{j\neq 0} e^{i\left(k_{{\bQp}=0}|\mathbf{R}_\parallel^j|-{\bQp}\cdot\mathbf{R}_\parallel^j\right)}
\left[ \left(\frac{k_{{\bQp}=0}^2}{|\mathbf{R}_\parallel^j|}+  \frac{i\, k_{{\bQp}=0}}{|\mathbf{R}_\parallel^j|^2}-\frac{1}{|\mathbf{R}_\parallel^j|^3}\right)\mathbb{1}\right. \nonumber\\
&-
\left. \frac{1}{|\mathbf{R}_\parallel^j|^3}  \left(k_{{\bQp}=0}^2+\frac{3i\,k_{{\bQp}=0}}{|\mathbf{R}_\parallel^j|}- \frac{3}{|\mathbf{R}_\parallel^j|^2} \right)\begin{pmatrix}
 \mathbf{R}_\parallel^j\otimes{\mathbf{R}_\parallel^j}  & 0 \\ & \\
0 & 0
\end{pmatrix}\right], \nonumber
\end{align}
accounts for particle interactions, with the $3\times3$ unity matrix $\mathbb{1}$.
The integer $j$ indexes the nanoparticles at the in-plane lattice vector positions of the 2D PC. Therefore, the form factor depends on the lattice structure and lattice constant $a$.
If the momentum ${\bQp}$ equals $\mathbf{0}$ or a reciprocal lattice vector ${\bGp}$, the form factor becomes diagonal with the entries~\cite{auguie_collective_2008},
\begin{align}
&\mathbcal{F}_{{\bQp}=\mathbf{0},ii}(\omega)=
\mathbcal{F}_{{\bGp},ii}(\omega)=\label{eq: PC Strukturefaktor}\sum_{j\neq 0} e^{ik_{{\Qp}=0}|\mathbf{R}_\parallel^j|}\,\, \times\\
&\times\left(\frac{(1-ik_{{\Qp}=0}|\mathbf{R}_\parallel^j|)(3X^j_i-1)}{|\mathbf{R}_\parallel^j|^3}+\frac{k_{{\bQp}=0}^2(1-X^j_i)}{|\mathbf{R}_\parallel^j|}\right)\nonumber. 
\end{align}
The angle dependency of the summands is given by $X_i^j=(\cos^2\theta^j,\sin^2\theta^j,0)$ for the entries $i=x,y,z$, respectively, with the polar angle $\theta_j$ of $\mathbf{R}_\parallel^j$.
In the numerical evaluation of the form factor, Eq.~\eqref{eq: PC Strukturefaktor}, the sum is only slowly converging, leading to spurious oscillations. This unphysical result can be circumvented using Ewald’s onefold integral transform~\cite{poppe_ir_1991}, yielding a fast converging expression for the form factor.

The Umklapp processes in Eq.~\eqref{eq: plasmon polarization} are considered in the near-field of the sample, but not in the far-field $(\vert z-z_\text{pl}\vert \rightarrow \infty)$.
This is appropriate since dipole densities with in-plane momentum $\bQp$ provide purely evanescent electric fields if the wave number $k_{\bQp}$ is imaginary, cf.~Eq.~\eqref{eq: Green's function, omega dependent}.
It follows that the ${\bGp}=0$ summand is the only propagating contribution if
\begin{align}
\Re(k_{\bGp \neq 0}) = 0 \Leftrightarrow
\Re(k_{\bGp^\text{min} \neq 0}) = 0.\label{eq: dense PC approximation (pre)}
\end{align}
The absolute value of the minimal non-trivial reciprocal lattice vector is $\vert \bGp^\text{min} \vert = 2\pi/a$ with the lattice constant $a$, as depicted in Fig.~\ref{fig: sketch hybrid}.
Consequently, Eq.~\eqref{eq: dense PC approximation (pre)} can be rewritten as a condition for the lattice constant $a$ compared to the wavelength $\lambda$ of the incoming light,
\begin{align}
a<\frac{\lambda}{\sqrt{\epsilon}}.\label{eq: dense PC approximation}
\end{align}
If we consider the sum over reciprocal lattice vectors in Eq.~\eqref{eq: plasmon polarization} as a diffraction phenomenon, Eq.~\eqref{eq: dense PC approximation} states that the lattice constant $a$ is small enough that the diffraction pattern in the far-field only exhibits the main maximum and not any additional side lobes. However, for the self-consistent solution, we have to consider all Umklapp processes in Eq.~\eqref{eq: plasmon polarization}, since both propagating and evanescent electric fields couple to the TMDC excitons which are located in the near-field of the PC.

\subsection{Plexcitons\label{sec: plexciton}}
In the following, the electric field, Eq.~\ref{eq: full electric field start}, and the dipole densities, Eq.~\ref{eq: definition, TMDC polarization} and \ref{eq: plasmon polarization}, are solved self-consistently to provide the near-field and far-field response of 2D TMDC-PC hybrids to an initially incident plane wave. 
To solve the set of equations, we insert the dipole densities of the individual layers, Eqs.~\eqref{eq: matrix definition, TMDC polarization} and \eqref{eq: plasmon polarization}, into the solution of the electric field, Eq.~\eqref{eq: full electric field start}, and altogether in the excitonic Bloch equation, Eq.~\eqref{eq:excitonicBlochEquation}.
From this procedure, we obtain a Bloch equation for the excitonic transition $\mathbf{p}_{\bQp}$,
\begin{widetext}
\begin{align}
\left( \hbar\omega-\mathcal{E}_{\bQp}+i\gamma  \right) \mathbf{p}_{\bQp}(\omega) = 
- \mathbcal{d}^* \cdot & \left( \mathbf{E}^0_{{\bQp}}(z_\text{ex},\omega)\right.\nonumber\\
&+\mathbcal{G}_{{\bQp}}(z_\text{ex},z_\text{ex},\omega)\cdot 
\mathbcal{d} \cdot \mathbf{p}_{{\bQp}}(\omega) \nonumber\\
&+\mathbcal{G}_{{\bQp}}(z_\text{ex},z_\text{pl},\omega)\cdot \boldsymbol{\alpha}^*_{\bQp}(\omega)\cdot\hspace{-0.4cm} \sum\limits_{\scriptscriptstyle\mathbf{\Qp^\prime}={\bQp}+{\bGp}}\hspace{-0.4cm}\mathbf{E}^0_{\bQp^\prime}(z_\text{pl},\omega) \nonumber\\
&+ \left.\mathbcal{G}_{{\bQp}}(z_\text{ex},z_\text{pl},\omega)\cdot\boldsymbol{\alpha}^*_{\bQp}(\omega)\cdot\hspace{-0.4cm}\sum\limits_{\scriptscriptstyle\mathbf{\bQp^\prime}={\bQp}+{\bGp}}\hspace{-0.4cm}\mathbcal{G}_{\bQp^\prime}(z_{\text{pl}},z_\text{ex},\omega)\cdot \mathbcal{d} \cdot \mathbf{p}_{\bQp^\prime}(\omega)\right),
\label{eq:MaxwellBlochEquation1}
\end{align}
\end{widetext}
where the left-hand side describes the free propagation of the excitons, cf.~Eq.~\eqref{eq:excitonicBlochEquation}, and on the right-hand side the self-consistent electric field occurs as a function of the dipole densities.

The first term appearing on the right-hand side of \eqref{eq:MaxwellBlochEquation1}, after the left parentheses,
$\mathbf{E}^0_{{\bQp}}(z_\text{ex},\omega)$, accounts for the undisturbed incoming electric field contribution at the TMDC position;
the second term carries the renormalization of the exciton line due to radiative self-interactions;
the third term accounts for the incoming field $\mathbf{E}^0_{{\bQp}}(z_\text{pl},\omega)$ scattered at the PC;
and the fourth (final) term, is the electric field generated by the excitons and back-scattered from the PC. This term is a radiative near-field interaction that couples excitonic transitions with momenta ${\bQp}$ to those ${\bQp^\prime}$ which are shifted by reciprocal lattice vectors $\mathbf{g}_\parallel$ of the PC compared to the excitonic momentum on the left-hand side.
This coupling results from the discretized translational invariance and will later be interpreted as a binding potential for originally spatially extended excitons.
The terms two and four give rise to an intervalley coupling in Eq.~\eqref{eq:MaxwellBlochEquation1}, since their prefactor matrices are non-diagonal in the valley index.

Since for the solution of Eq.~\eqref{eq:MaxwellBlochEquation1}, we need to consider the observables near the sample, i.e., in a near-field coupling, we apply a quasi-static approximation in Eq.~\eqref{eq:dyadic Green's function}. We neglect propagation effects by setting $\omega$ to zero and otherwise only consider a parametric $\omega$-dependency, e.g., in the permittivity $\varepsilon_\text{Au}(\omega)$, the polarization densities or the electric field.
This approximation is valid as long as~\cite{jackson_klassische_2014}
\begin{align}
    \lambda\gg  \vert z_\text{pl}-z_\text{ex}\vert,
\end{align}
where the wavelength $\lambda$ of the incoming light is compared to the distance between the two lattices.
Thus, we obtain the simplified (quasi-static) scalar Green's function
\begin{align}
    G_{\Qp}(z,z')&=\frac{-1}{2\Qp}e^{-\Qp|z-z'|}\label{eq: Greens function Quasi-static},
\end{align}
which shows an exponential decrease that depends on the distance between the source of the electric field at $z^\prime$ and the observer with position $z$. We indicate the quasi-static Green's function by dropping the dependence on the frequency $\omega$.
However, the Green's dyadic, Eq.~\eqref{eq:dyadic Green's function}, shows that the quasi-static approximation corresponds to neglecting $\omega^2$ compared to the in-plane momentum $\Qp^2$ which is only reasonable if
\begin{align}
    \frac{\omega^2}{c^2} < \frac{\Qp^2}{\epsilon}.\label{eq: quasi-static condition}
\end{align}
In the following, we refer to electric fields and dipole densities that fulfill the condition in Eq.~\eqref{eq: quasi-static condition} as \textit{outside the light-cone}. No propagating solution of the wave equations is possible, since the electric field is damped exponentially with the distance, cf.~Eqs.~\eqref{eq: Green's function, omega dependent} and \eqref{eq: Greens function Quasi-static}. In contrast, in-plane momenta that do not fulfill the condition are \textit{inside the light-cone}.
The condition \eqref{eq: quasi-static condition} is only valid for excitonic momenta ${\bQp}$ belonging to originally momentum-dark excitonic transitions.
However, an inspection of Eqs.~\eqref{eq: full electric field start} and \eqref{eq: plasmon polarization} shows that, for a perpendicular plane wave excitation, cf. Eq.~\eqref{eq: plane wave E0},
the plasmonic dipole density is only non-zero for in-plane momenta equal to a reciprocal lattice vector $\bGp$ of the PC. Since only the zeroth order scattering momentum $\bGp=0$ is a propagating solution (inside the light cone), it is allowed to apply the quasi-static approximation for all momenta ${\bQp}$ of the excitonic transition, except for ${\bQp}=\mathbf{0}$.
Therefore, we split Eq.~\eqref{eq:MaxwellBlochEquation1} into a near-field Bloch equation ($\Qp>0$) and a radiative contribution ($\Qp=0$) and discuss them separately in the following (Sec.~\ref{sec: near field exciton-plasmon interaction}, \ref{radiative exciton-plasmon interaction}, respectively).
In particular, by inserting the Green's function for a homogeneous environment permittivity in quasi-static approximation, Eq.~\eqref{eq: Greens function Quasi-static}, one observes that the right-hand side of Eq.~\eqref{eq:MaxwellBlochEquation1} is proportional to the in-plane momentum. The momentum-dark excitonic transition $\mathbf{p}_{{\bQp}\neq \mathbf{0}}(\omega)$ (Sec.~\ref{sec: near field exciton-plasmon interaction}) does not couple to the radiative excitonic transition $\mathbf{p}_{{\bQp}= 0}(\omega)$ (Sec.~\ref{radiative exciton-plasmon interaction}) and can be solved independently.

\subsubsection{Near-Field Exciton-Plasmon Interaction\label{sec: near field exciton-plasmon interaction}}
The resulting Bloch equations for $\Qp\neq 0$ are identical to Eq.~\eqref{eq:MaxwellBlochEquation1}, with the quasi-static approximation applied to every Green's dyadic.
To account for the circular dichroism of the TMDC excitons,
we choose a circular polarized basis $(\mathbf{e}^+,\mathbf{e}^-,\mathbf{e}^z)$.
Any quantity in Cartesian coordinates is transformed to the circular polarized basis by multiplying with the unitary (change-of-basis) matrix $\mathbcal{T}^{\text{circ}}$,
\begin{align}
\begin{pmatrix}
E^{+}\\
E^{-}\\
E^z
\end{pmatrix}
=
\frac{1}{\sqrt{2}}
\begin{pmatrix}
1 & -i & 0\\
1 & i & 0\\
0 & 0 & \sqrt{2}
\end{pmatrix}
\cdot
\begin{pmatrix}
E^{x}\\
E^{y}\\
E^{z}
\end{pmatrix}
\equiv
\mathbcal{T}^{\text{circ}}
\cdot
\begin{pmatrix}
E^{x}\\
E^{y}\\
E^{z}
\end{pmatrix}
.
\end{align}
In the quasi-static limit, a diagonalization of  Eq.~\eqref{eq:MaxwellBlochEquation1} in circular polarized basis corresponding to the valleys $K^+$ and $K^-$ is given by a transformation $T_\phi$, Eq.~\eqref{eq: T Phi}, that depend on the polar angle $\phi$ of the in-plane momentum ${\bQp}$,
\begin{align}
\begin{pmatrix}
p^U_{\bQp}\\
p^V_{\bQp}\\
p^z_{\bQp}
\end{pmatrix}
=
\frac{1}{\sqrt{2}}
\begin{pmatrix}
-e^{i\phi} & e^{-i\phi} & 0\\
e^{i\phi} & e^{-i\phi} & 0\\
0 & 0 & \sqrt{2}
\end{pmatrix}
\cdot
\begin{pmatrix}
p^{+}_{\bQp}\\
p^{-}_{\bQp}\\
p^{z}_{\bQp}
\end{pmatrix}
\equiv
\mathbcal{T}_\phi
\cdot
\begin{pmatrix}
p^{+}_{\bQp}\\
p^{-}_{\bQp}\\
p^{z}_{\bQp}
\end{pmatrix}
.
\label{eq: T Phi}
\end{align}
The transformation $\mathbcal{T}_\phi$ corresponds to an in-plane rotation of the momentum space, orienting the momentum ${\bQp}$ along the $V$-axis.
We find that the transformation into the $(\mathbf{e}^U,\mathbf{e}^V,\mathbf{e}^z)$ basis is the diagonalization of the in-plane contribution of the quasi-static Green's dyadic,
\begin{align*}
\mathbcal{G}^{U/V}_{{\bQp}}(z,z') =& \frac{1}{\varepsilon_0\varepsilon}
\begin{pmatrix}
0 & 0 & 0 \\
0 & \Qp^2 & i \Qp \partial_{z'} \\
0 & i \Qp \partial_{z'} & - \partial_{z'}^2
\end{pmatrix}
G_{\Qp}(z,z').
\end{align*}
Applying the rotation to the Bloch equations \eqref{eq:MaxwellBlochEquation1} finally allows a diagonalization process,
\begin{align}
&\left( \hbar\omega - \mathcal{E}_{\bQp}+i\gamma  \right) p^U_{{\bQp}\neq \mathbf{0}}(\omega)
= 0,\label{eq:MaxwellBlochEquationparabolic}\\
&\left( \hbar\omega- \mathcal{E}^V_{\bQp}
+i\gamma\right) p^V_{{\bQp}\neq \mathbf{0}}(\omega) +
\hspace*{-0.4cm}\sum\limits_{\scriptscriptstyle\mathbf{\Qp^\prime}={\bQp}+{\bGp}}\hspace*{-0.4cm}
C^{\text{ex} \leftrightarrow \text{pl}}_{{\bQp}\mathbf{\Qp^\prime}\omega}\, p^V_{\bQp^\prime}(\omega) \nonumber \\
&= -\frac{ d^*}{\epsilon_0 \epsilon}
\mathbf{S}^{*\text{ex}\leftarrow\text{pl}}_{{\bQp}}(\omega) \hspace{-0.01cm}\cdot \hspace{-0.3cm}
\sum\limits_{\scriptscriptstyle\mathbf{\Qp^\prime}={\bQp}+{\bGp}} \hspace{-0.3cm} \mathbf{E}^0_{\bQp^\prime}(z_\text{pl},\omega),
\label{eq:MaxwellBlochEquationconic}
\end{align}
where Eq.~\eqref{eq:MaxwellBlochEquationparabolic} possesses the undisturbed, parabolic ($U$-shaped) exciton dispersion for $p^U_{\bQp}$.
In contrast, we find
\begin{align}
    &\mathcal{E}^V_{\bQp} = 
    E^\text{ex}
    +\frac{\hbar^2 \Qp^2}{2M}
    -\frac{ |d|^2}{\epsilon_0 \epsilon} \Qp^2 \, G_{\Qp}^{\text{ex}- \text{ex}}
    \approx E^\text{ex}+\beta \Qp,\\
    &\text{with} \hspace{0.5cm}\beta = \frac{ |d|^2}{2\epsilon_0 \epsilon}\nonumber,
\end{align}
yielding a conic ($V$-shaped) dispersion for $p^V_{\bQp}$ similar to
results reported in Ref.~\cite{qiu_nonanalyticity_2015}. The excitonic self-interaction term, $\Qp^2\, G_{\Qp}^{\text{ex}- \text{ex}}$, dominates the kinetic energy contribution for small $\Qp$. 
The optical sources of $p^V_{\bQp}$, cf.~Eq.~\eqref{eq:MaxwellBlochEquationconic}, occur on the right.
Obviously, the near-field interaction of exciton and plasmon is fully encoded in $p^V_{\bQp}$.
The interaction with the PC is denoted by the contributions $(\text{ex}\hspace{-0.1cm}\leftrightarrow\hspace{-0.1cm}\text{pl})$ and provides a momentum and frequency dependent PC-induced excitonic self-interaction $C^{\text{ex} \leftrightarrow \text{pl}}_{{\bQp}\mathbf{\Qp^\prime}\omega}$, namely, a plasmon mediated effective exciton-exciton interaction. For a concise notation without loss of information, we indicate the $z$-dependencies of the prefactor Green's functions in their superscripts. 
On the right-hand side of Eq.~\eqref{eq:MaxwellBlochEquationconic}, the existence of an incoming field $\mathbf{E}^0_{\bQp}$, that was scattered at the PC, results in the excitation of the conic excitonic transition.
The interaction with the PC provides a coupling to in-plane momenta shifted by a reciprocal lattice vector $\mathbf{g}_\parallel$ of the PC due to Umklapp processes depending on the PC lattice structure.
The PC-induced excitonic self-interaction reads
\begin{align}
C^{\text{ex} \leftrightarrow \text{pl}}_{{\bQp}\mathbf{\Qp^\prime}\omega}=  &
\left(\cos\phi\cos\phi' \alpha^*_{xx}(\omega)
+   \sin\phi\cos\phi' \alpha^*_{yx}(\omega) \right. \nonumber \\ 
& + \cos\phi\sin\phi' \alpha^*_{xy}(\omega) + \sin\phi\sin\phi' \alpha^*_{yy}(\omega)  \nonumber \\
&+\left.  \alpha^*_{zz}(\omega) \right)  \frac{ |d|^2}{(\epsilon_0 \epsilon)^2}  \Qp^2 {\Qp^\prime}^2 G_{\Qp}^{\text{ex-pl}} G_{\Qp^\prime}^{\text{pl-ex}}  \label{eq:auxiliaryV},
\end{align}
and the source term that arises from the scattering of the incoming field at the PC becomes
\begin{align}
\mathbf{S}^{\text{ex}\leftarrow\text{pl}}_{{\bQp}}(\omega)
=&\,
\Qp^2 G_{\Qp}^{\text{ex}-\text{pl}}
\begin{pmatrix}
\alpha^*_{xx}(\omega)\cos\phi+\alpha^*_{yx}(\omega)\sin\phi \\
\alpha^*_{xy}(\omega)\cos\phi+\alpha^*_{yy}(\omega)\sin\phi \\
i \alpha^*_{zz}(\omega) \text{sgn}(z_{ex}-z_{pl})
\end{pmatrix}
\label{eq:auxiliaryS},
\end{align}
if we express  $\mathbf{E}^0_{\bQp^\prime}(z_\text{pl},\omega)$ in Cartesian coordinates.
To achieve these expressions, derivatives of the Green's function with respect to $z^\prime$ have been carried out, cf.~Eq.~\eqref{eq:dyadic Green's function}.

Since we were able to decouple the valleys of the excitonic transition $\{p^{+}_{\bQp}\hspace{-0.1cm},p^{-}_{\bQp}\}$ in Eq.~\eqref{eq:MaxwellBlochEquation1} into a conic (optically driven $p^V_{\bQp}$, Eq.~\eqref{eq:MaxwellBlochEquationconic}) and an undisturbed parabolic ($p^U_{\bQp}$, Eq.~\eqref{eq:MaxwellBlochEquationparabolic}) contribution, we can now solve them independently.
The parabolic Bloch equation for $p^U_{\bQp}$ immediately yields
\begin{align}
 p^U_\mathbf{\Qp\neq 0}(\omega)
=0 .
\end{align}
In contrast, in Eq.~\eqref{eq:MaxwellBlochEquationconic}, the conic Bloch equation for $p^V_{\bQp}$ contains a coupling of excitons with different center-of-mass momenta, mediated by the PC.
We project the excitonic transition $p^V_{\bQp}$ onto a set of eigenstates of the conic dispersion,
\begin{align}
p^V_\mathbf{\Qp\neq 0}(\omega)  = \sum_{\lambda} v^{R\lambda}_{\bQp}(\omega) p^{\lambda}(\omega).\label{eq:pHprojectionRule}
\end{align}
The corresponding symmetric, non-Hermitian eigenvalue problem reads similar to Ref.~\cite{salzwedel_spatial_2023},
\begin{align}
 \mathcal{E}^V_{\bQp}
 v^{R\lambda}_{{\bQp}}(\omega)    
-\hspace{-0.5cm} \sum\limits_{\scriptscriptstyle\mathbf{\Qp^\prime}={\bQp}+{\bGp}}\hspace{-0.4cm} 
C^{\text{ex} \leftrightarrow \text{pl}}_{{\bQp}\mathbf{\Qp^\prime}\omega}\, v^{R\lambda}_{\bQp^\prime}(\omega) 
=E^\lambda(\omega) v^{R\lambda}_{\bQp}(\omega)\label{eq:conicEVProblem}.
\end{align}
As a consequence of the non-Hermiticity we have to distinguish between left and right eigenstates $v^{L\lambda}_{\bQp},\, v^{R\lambda}_{\bQp}$.
To justify the projection onto these states, we verify the existence of solutions of Eq.~\eqref{eq:conicEVProblem} and their completeness numerically.
By applying a suitable normalization, we ensure orthonormality using
\begin{align}
 \sum_{\bGp} 
 \left(v^{L\lambda}_{{\bQp}+{\bGp}}(\omega)\right)^*
 \left(v^{R\lambda^\prime}_{{\bQp}+{\bGp}}(\omega) \right)
 =\delta_{\lambda\lambda^\prime}.\label{eq: orthogonality eigenstates}
\end{align}
With this expression, we project the conic excitonic transition in the Bloch equation \eqref{eq:MaxwellBlochEquationconic} on the new states according to Eq.~\eqref{eq:pHprojectionRule}. Substituting the momentum dependencies with the corresponding eigenvalue in Eq.~\eqref{eq:conicEVProblem} and taking advantage of the orthonormality condition between the eigenstates finally allows us to give a solution for the conic contribution of the excitonic transition
\begin{align}
   p^V_\mathbf{\Qp\neq 0}(\omega)&=
-d^*\hspace{-0.4cm}\sum\limits_{\lambda,\bQp^\prime={\bQp}+{\bGp}}\hspace{-0.4cm}
\frac{v^{R\lambda}_{{\bQp}}(\omega) v^{L\lambda^*}_{{\bQp}^\prime}(\omega)}{\hbar\omega-E^{\lambda}(\omega)+i\gamma  }\times\\
&\times\left(\frac{1}{\epsilon_0 \epsilon}
\mathbf{S}^{*\text{ex}\leftarrow\text{pl}}_{\bQp^\prime}(\omega) \cdot 
\sum\limits_{{\bGp}}  \mathbf{E}^0_{{\bQp}+{\bGp}}(z_\text{pl},\omega)\right).\nonumber
\end{align}
To connect the excitonic transition with observables, we insert it in the dipole density of the TMDC layer. The contribution arising from the conic part becomes
\begin{align}
    & {P^{2\text{D},\text{ex}}_{{\bQp}\neq \mathbf{0}}\hspace{0.05cm}}^{\pm}(\omega) = d\frac{e^{\mp i\phi}}{\sqrt{2}} \sum\limits_{\lambda} v^{R\lambda}_{{\bQp}}(\omega) p^\lambda(\omega).
\end{align}
The full dipole density stemming from momentum-dark excitons trapped in the plasmonic potential is finally given by
\begin{align}
{P^{2\text{D},\text{ex}}_{{\bQp}={\bGp}\neq \mathbf{0}}\hspace{-0.3cm}}^{\pm}\hspace{0.05cm}(\omega)&=
-\frac{\vert d\vert ^2}{\sqrt{2}{\epsilon\epsilon_0}} e^{\mp i\phi}\times
\label{eq: Q neq 0 polarization}\\
\times\sum\limits_{\lambda{\bGp}}& \frac{v^{R\lambda}_{{\bQp}}(\omega)v^{L\lambda^*}_{{\bGp}}(\omega)}{\hbar\omega - E^\lambda + i\gamma}\mathbf{S}^{*\text{ex}\leftarrow\text{pl}}_{\bGp^\prime}(\omega) \cdot \mathbf{\tilde{E}}^0_{ {\bGp}}(z_{\text{pl}},\omega).\nonumber
\end{align}
For momenta that do not correspond to the wave number of the incoming field shifted by a reciprocal lattice vector, the TMDC polarization vanishes due to the absence of a suitable excitation.

\subsubsection{Radiative Exciton-Plasmon Interaction\label{radiative exciton-plasmon interaction}}
In the previous section, the excitonic dipole density outside the light-cone was derived. However, to determine the far-field response, we need an expression for the dipole density within the light cone which is a source for propagating solutions of the electric field.
Therefore, we evaluate the Bloch equation \eqref{eq:MaxwellBlochEquation1} for $\Qp=0$, i.e., the radiative Bloch equations,
\begin{align}
&\left( \hbar\omega-E^\text{ex}+i\gamma  \right) \mathbf{p}_{{\bQp}=\mathbf{0}}\nonumber\\
= 
&-\biggl(
\mathbcal{d}^*\cdot\mathbcal{G}(z_\text{ex},z_\text{pl},\omega)\cdot \alpha^*_{{\bQp}=\mathbf{0}}(\omega)\cdot \mathbf{E}^0_{{\bQp}=\mathbf{0}}(z_\text{pl},\omega)\nonumber\\
+&\mathbcal{d}^*\cdot\mathbcal{G}(z_\text{ex},z_\text{pl},\omega)\cdot\boldsymbol{\alpha}^*_{{\bQp}=\mathbf{0}}(\omega)\cdot\hspace{-0.4cm}\sum\limits_{\scriptscriptstyle\mathbf{\Qp^\prime}={\bQp}+\mathbf{g}_\parallel}\hspace{-0.3cm}\mathbcal{G}_{\bQp^\prime}(z_{\text{pl}},z_\text{ex})\cdot \mathbcal{d} \cdot \, \mathbf{p}_{\bQp^\prime}\nonumber \\
+& \mathbcal{d}^*\cdot\mathbcal{G}(z_\text{ex},z_\text{pl},\omega)\cdot\boldsymbol{\alpha}^*_{{\bQp}=\mathbf{0}}(\omega)
\cdot
\mathbcal{G}(z_{\text{pl}},z_\text{ex},\omega)\cdot \mathbcal{d} \cdot \, \mathbf{p}_{{\bQp}=\mathbf{0}}\nonumber \\
+&\mathbcal{d}^*\cdot\mathbcal{G}(z_\text{ex},z_\text{ex},\omega)\cdot 
\mathbcal{d} \cdot \, \mathbf{p}_{{\bQp}=\mathbf{0}} \biggr) \nonumber\\
&- \mathbcal{d}^* \cdot\mathbf{E}^0_{{\bQp}=\mathbf{0}}(z_\text{ex},\omega).\label{eq:MaxwellBlochEquation Q=0}
\end{align}
For clarity, we suppress the index of the vanishing in-plane momentum ${\bQp}=\mathbf{0}$ in the notation for the radiative Green's dyadic.
The previous provides us with all the necessary tools to solve this equation. Simplifying the individual terms on the right-hand side is demanding but straight-forward by explicitly evaluating the matrix products between the dipole elements, Eq.~\eqref{eq: definition dipole tensor}, the Green's dyadic, Eq.~\eqref{eq:dyadic Green's function} and the PC polarizability $\boldsymbol{\alpha}^*_{\bQp}(\omega)$. Finally, we collect all summands containing $\mathbf{p}_{{\bQp}=\mathbf{0}}$ on the left-hand side and multiply with the inverse of its prefactor. This procedure yields
\begin{widetext}
\begin{align}
&\mathbf{p}_{{\bQp}=\mathbf{0}}
= \left[ \left(\hbar\omega-E^\text{ex}+i\gamma
-\vert d \vert^2 \frac{\omega^2}{\varepsilon_0 c^2}
G_{ \Qp=0,\omega}^{\text{ex-ex}} \right)\mathbb{1}
+\vert d \vert^2\,\left(\frac{\omega^2}{\varepsilon_0 c^2}\right)^2\,G_{ \Qp=0,\omega}^{\text{ex-pl}}\,G_{ \Qp=0,\omega}^{\text{pl-ex}} 
\boldsymbol{\alpha}^*_{{\bQp}=\mathbf{0}}(\omega)
\right]^{-1}\cdot\label{eq: radiative excitonic transition}\\
&\cdot\hspace{-0.15cm}\left[ 
- d^*\mathbf{E}^0_{{\bQp}=\mathbf{0}}(z_\text{ex},\omega)
+d^*\frac{\omega^2}{\varepsilon_0 c^2}G_{ \Qp=0,\omega}^{\text{ex-pl}}
\boldsymbol{\alpha}^*_{{\bQp}=\mathbf{0}}\hspace{-0.2cm}
\cdot
\mathbf{E}^0_{{\bQp}=\mathbf{0}}(z_\text{pl},\omega)
+ \vert d \vert^2 \frac{\omega^2}{\varepsilon_0 c^2}
\sum\limits_{\scriptscriptstyle\mathbf{\Qp^\prime}=\mathbf{g}_\parallel}
\frac{{\Qp^\prime}^2}{\varepsilon_0\varepsilon}
G_{ \Qp=0,\omega}^{\text{ex-pl}} G_{\Qp^\prime,\omega=0}^{\text{pl-ex}}
\boldsymbol{\alpha}^*_{{\bQp}=\mathbf{0}}
\hspace{-0.2cm}
\cdot
\mathbf{e}_{\bQp^\prime}
\,
p^{V}_{\bQp^\prime}\right],\nonumber
\end{align}
\end{widetext}

with the unity vector in polar coordinates
\begin{align*}
\mathbf{e}_{\bQp^\prime}=
    \begin{pmatrix}
    \cos{\phi^\prime}\\
    \sin{\phi^\prime}\\
    0
\end{pmatrix},
\end{align*}
when we express the excitonic transition $\mathbf{p}_{{\bQp}=\mathbf{0}}$ and the PC polarizability $\boldsymbol{\alpha}^*_{{\bQp}}$ in a linear polarized basis.

The denominator of Eq.~\eqref{eq: radiative excitonic transition} corresponds to harmonic oscillators at frequency $\omega$ with renormalized eigenenergies
\begin{align}
   E^\text{ex}\,\mathbb{1} - \vert d \vert^2\,\left(\frac{\omega^2}{\varepsilon_0 c^2}G_{ \Qp=0,\omega}^{\text{ex-pl}}\right)^2\,
\re (\boldsymbol{\alpha}^*_{{\bQp}=\mathbf{0}}(\omega)),
\end{align}
and the damping
\begin{align*}
\gamma\,\mathbb{1}&+i\vert d \vert^2 \frac{\omega^2}{\varepsilon_0 c^2}
G_{ \Qp=0,\omega}^{\text{ex-ex}}\,\mathbb{1}\\
&+ \vert d \vert^2\left(\frac{\omega^2}{\varepsilon_0 c^2}G_{ \Qp=0,\omega}^{\text{ex-pl}}\!\right)^2
\!\! \im  (\boldsymbol{\alpha}^*_{{\bQp}=\mathbf{0}}(\omega))
.
\end{align*}
The ex-ex term accounts for the radiative dephasing of the TMDC excitons whereas the last term governs radiative interference phenomena between the TMDC and the PC. For a resonant TMDC-PC interaction ($E^\text{ex}=E^\text{pl}$), the real part of the PC polarizability can be neglected $\re (\boldsymbol{\alpha}^*_{\bQp}(\hbar^{-1}E^\text{pl}))\approx 0$. This yields an unchanged excitonic eigenenergy and the coupling to the PC mainly modifies the damping.
The numerator includes from left to right: the direct excitation of excitonic transitions via the incident electric field, the plasmon mediated excitation from the incoming field that was first scattered at the PC, and finally the plasmon-mediated influence of momentum-dark excitonic transitions.
From the derived excitonic transition, we deduce the dipole density via Eq.~\eqref{eq: matrix definition, TMDC polarization}.

\subsubsection{Scattered Electric Field}
To give explicit results for the electric field emitted by the TMDC layer $\mathbf{E}_{{\bQp}}^\text{ex}$, we multiply the dipole density, that was derived in the previous section, with the Green's dyadic. We find the momentum-dark dipole density to be an eigenstate of the quasi-static Green's dyadic, cf.~Eq.~\eqref{eq:dyadic Green's function}. Therefore, we find
\begin{align}
\mathbf{E}^{\text{ex}}_{{\bQp}\neq \mathbf{0}}(z,\omega) &= 
\frac{\Qp^2}{\epsilon\epsilon_0}
 G_{{\Qp}}(z,z_{\text{ex}},\omega=0)\,
\mathbf{P}^{2\text{D},\text{ex}}_{{\bQp}\neq \mathbf{0}}\label{eq: TMDC electric field final Q neq 0}\\
\mathbf{E}^{\text{ex}}_{{\bQp}=\mathbf{0}}(z,\omega) &=-
 \frac{\omega^2}{\epsilon_0 c^2} G_{{\Qp}=0}(z,z_{\text{ex}},\omega)\, \mathbf{P}^{2\text{D},\text{ex}}_{{\bQp}=\mathbf{0}}. \label{eq: TMDC electric field final Q eq 0}
\end{align}
In fact, also a $z$-component of the electric near-field $E^{\text{ex}^{z}}_{{\bQp}\neq \mathbf{0}}(z,\omega)$ occurs. However, it only causes a non-vanishing $z$-component of the PC polarization since $\boldsymbol{\alpha}^*_{{\bQp}}$ is assumed to be diagonal regarding its $z$-entries.
This $z$-polarization of the PC is not transferred to the far-field for ${\bQp}=\mathbf{0}$.
To obtain the contribution of the PC to the far-field, $\mathbf{E}_{{\bQp}=\mathbf{0}}^\text{pl}$, we deduce the dipole density of the 2D PC, according to Eq.~\eqref{eq: plasmon polarization}. It is
\begin{align}
\mathbf{P}^{\text{2D},\text{pl}}_{{\bQp}}(\omega)
	&= \frac{\boldsymbol{\alpha}^*_{{\bQp}}(\omega)}{A_{\text{UC}}^{\text{pl}}}\cdot\label{eq: final plasmon polarization}
 \\ &
 \sum\limits_{{\bGp}}\left(\mathbf{E}^0_{{{\bQp}+{\bGp}}}(z_{\text{pl}},\omega) + \mathbf{E}_{{\bQp}+{\bGp}}^{\text{ex}}(z_{\text{pl}},\omega)\right).\nonumber
\end{align}
As already justified above for dense PCs, cf.~Eq.~\eqref{eq: dense PC approximation}, there is no propagating solution of Maxwell's equations with an in-plane wavenumber equal to a non-trivial reciprocal lattice vector. Therefore, we may drop the sum over ${\bGp}\neq 0$ for the exciting field.
As for the radiative TMDC contribution, the dyadic Green's function has to contain the full time dependency.
Furthermore, since $z\neq z^\prime$, the $z$-entries of the Green's dyadic vanish. The in-plane far-field plasmon contribution reads
\begin{align}
&{\mathbf{E}}^\text{pl}_{{{\bQp}=\mathbf{0}}}(z,\omega) =
\frac{-\omega^2}{\epsilon_0 c^2}
G_{\Qp=0}(z,z_{\text{pl}},\omega)\,
\boldsymbol{\alpha}^*_{{\bQp}=\mathbf{0}}(\omega)\cdot\label{eq: PC electric field final}\\	&\cdot\left(\mathbf{E}^0_{{{\bQp}}=\mathbf{0}}(z_{\text{pl}},\omega)
	+
	\mathbf{E}_{{\bQp}=\mathbf{0}}^{\text{ex}}(z_{\text{pl}},\omega)
	+
	\sum\limits_{\bGp\neq \mathbf{0}}
	\mathbf{E}_{\bGp\neq \mathbf{0}}^{\text{ex}}(z_{\text{pl}},\omega)\right).\nonumber
\end{align}
Adding all contributions of the electric field according to Eq.~\eqref{eq: full electric field start}, results in 
\begin{equation}
\mathbf{E}_{{\bQp}=\mathbf{0}}(z,\omega)
=\mathbf{E}_{{\bQp}=\mathbf{0}}^{\text{ex}}(z,\omega) +
\mathbf{E}_{{\bQp}=\mathbf{0}}^{\text{pl}}(z,\omega)
+\mathbf{E}^0_{{\bQp}=\mathbf{0}}(z,\omega) 
\label{eq: full electric far field final}.
\end{equation}

\subsubsection{Overview of the Theory\label{sec: theory explanation}}

\begin{figure}[t]
 \begin{center}
 \includegraphics[width=\linewidth, trim={0 2cm 0 0}]{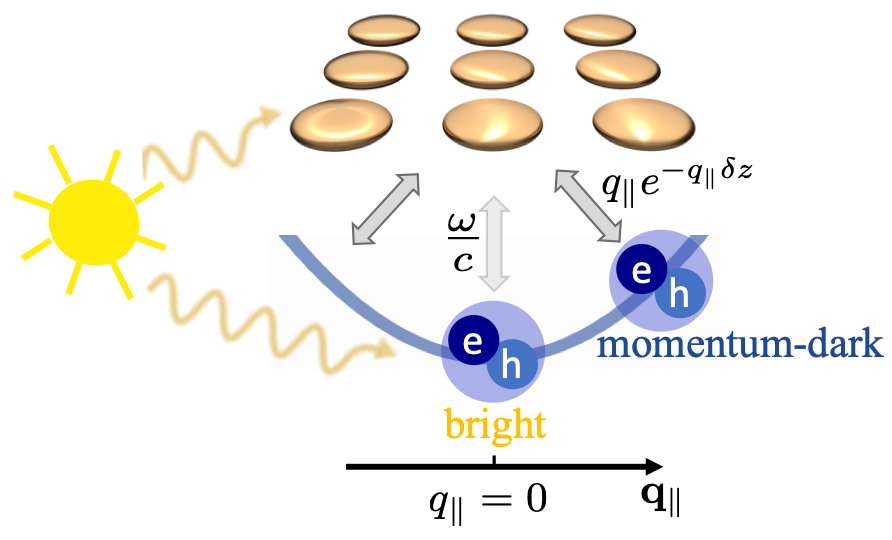}
 \end{center}
 \caption{\textbf{Schematic of the theory:} the PC couples to the bright and momentum-dark TMDC excitons which are illustrated on the parabolic dispersion curve. The sun represents the far-field excitation. The gray arrows account for the exciton-plasmon coupling, with the indicated coupling strengths.}
 \label{fig: scheme theory}
\end{figure}

With the theoretical framework now fully established, our attention turns to shedding some light on the physical meaning of the derived equations. Figure~\ref{fig: scheme theory} provides a schematic of the theory.

We excite the TMDC-PC hybrid via far-field illumination $\mathbf{E}^0_{\bQp}$, Eq.~\eqref{eq: plane wave E0}, represented by the sun-like schematic in Fig.~\ref{fig: scheme theory}. It directly acts on the PC plasmons, Eq.~\eqref{eq: final plasmon polarization}, and the bright excitons $\mathbf{p}_{{\bQp}=\mathbf{0}}$, Eq.~\eqref{eq: radiative excitonic transition}. The scattering respectively Umklapp processes of the incoming light at the PC provide access to momentum-dark excitonic transitions, Eq.~\eqref{eq:MaxwellBlochEquationconic}, via the plasmon-enhanced electric near-field. These interactions are typical dipole-dipole near-field interactions, proportional to $\Qp e^{-{\Qp}\delta z}$, which can be seen by explicitly inserting the quasi-static Green's function, Eq.~\eqref{eq: Greens function Quasi-static}, into the conic Bloch equation, Eq.~\eqref{eq:MaxwellBlochEquationconic}. The coupling between the plasmon and the bright exciton behaves qualitatively different with the coupling strength significantly reduced, proportional to ${\omega}/{c}$, cf. Eq.~\eqref{eq: radiative excitonic transition}. The implications of these different coupling mechanisms are numerically evaluated in Sec.~\ref{sec: results}.
Finally, we note that, in the limit $a\rightarrow\infty$, the developed theory is also applicable for a TMDC coupled to a single MNP without qualitatively changing the results.

\clearpage

\onecolumngrid

\begin{figure}[b]
 \begin{center}
    \includegraphics[width=\linewidth, trim={0 .4cm 0 0}]{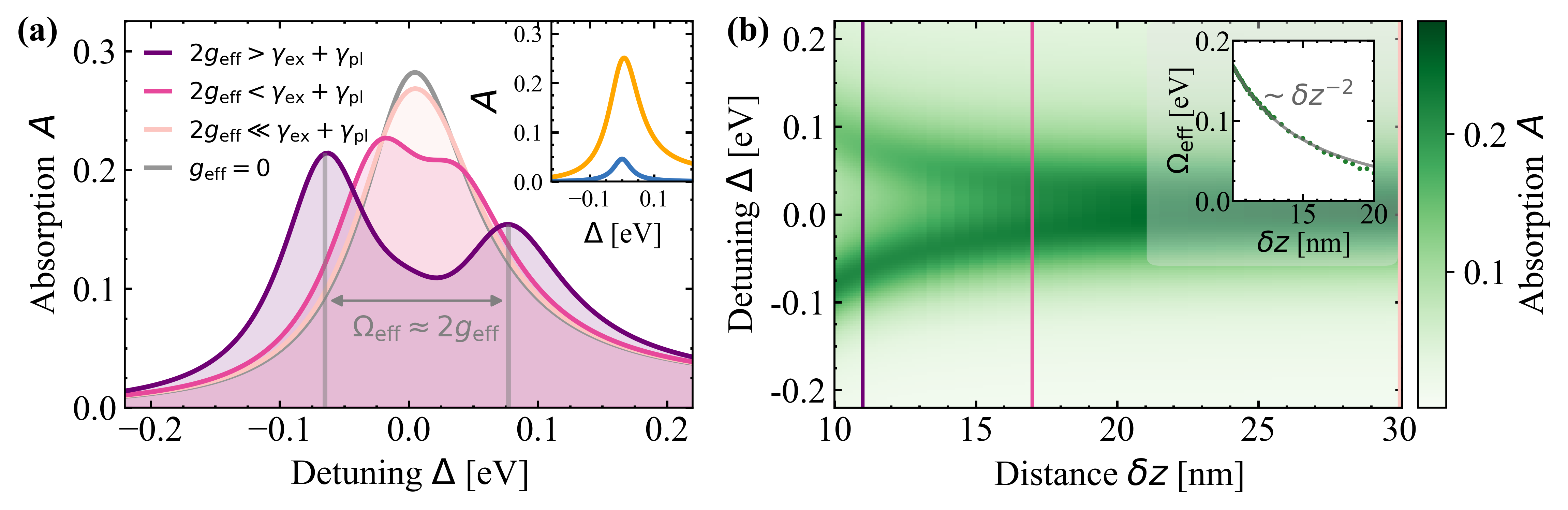}
 \end{center}
\caption{
\textbf{Absorption for different distances} $\delta z =\vert z_\text{pl} - z_\text{ex}\vert$ between TMDC and PC at room temperature, corresponding to a scan over an effective coupling strength $g_{\mathrm{eff}}$.
\textbf{(a)} Absorption spectra for $\delta z =[11,17,30]\,$nm belonging to different interaction regimes, from weak $(2g_{\mathrm{eff}}<\gamma_{\mathrm{ex}}+\gamma_{\mathrm{pl}})$ to strong $(2g_{\mathrm{eff}}>\gamma_{\mathrm{ex}}+\gamma_{\mathrm{pl}})$ coupling. For comparison, the artificially uncoupled case (gray), is the sum of single TMDC (blue) and PC (yellow) absorption in the inset.
\textbf{(b)} Color map of the absorption depending on the distance $\delta z$ and the Detuning $\Delta$. The vertical lines indicate the spectra shown in panel a).
\textbf{Inset:} Effective Rabi energy $\Omega_\text{eff}$ over $\delta z$. It reaches values significantly larger than $100\,$meV. The peak splitting extracted from the full theory is compared to a fit $b/(\delta z)^2$ with $b=17.24\,\text{nm}^2\text{eV}$. \\}
 \label{fig: Delta z scan}
\end{figure}
\twocolumngrid

\section{Numerical Results and Discussion\label{sec: results}}
We numerically evaluate the key equations \eqref{eq: PC electric field final},\,\eqref{eq: TMDC electric field final Q eq 0},\,\eqref{eq: TMDC electric field final Q neq 0},\,\eqref{eq: Q neq 0 polarization} and \eqref{eq:conicEVProblem} to calculate the absorption, Eq.~\eqref{eq: TRA 1}, of the hybrid structure.
If not stated differently, we perform calculations for a MoSe$_2$-PC stack with the geometry specified in Table \ref{table: hybrid geometry}. The given parameters yield a PC with its resonance energy at $E^\text{pl}=2.03\,$eV.
We choose the lattice constant $a$ in a way that the collective PC resonance sharpens the single MNP response that is governed by the bulk metal parameters and aspect ratio $r_z/r_x$.
Material specific parameters for gold and MoSe$_2$ are given in Appendix A.
\begin{table}[htb]
	\centering
	\caption{Geometry of TMDC-PC hybrid}
	\begin{tabular}{c|l||c|l}
	$r_x$ & $30\,$nm & $a$ & $10\, r_x$  \\\hline
    $r_y$ & $30\,$nm  & $\delta z$ & $11\,$nm  \\\hline
    $r_z$ & $10\,$nm  & $\varepsilon$ & $2.4$
	\end{tabular}
 \label{table: hybrid geometry}
\end{table}

\subsection{Coupling Regimes at Room Temperature}
To study different coupling regimes, the modification of  the layer distance $\delta z$ allows to directly access the strength of the near-field mediated interaction. Figure~\ref{fig: Delta z scan} shows the absorption of the hybrid over the detuning $\Delta = \hbar\omega-E^\text{pl}$ between the electric field and the plasmon for the resonant case for exciton and plasmon $(E^\text{ex}=E^\text{pl})$
for different exciton-plasmon distances $\delta z$. Modifying the distance $\delta z$ qualitatively changes the hybrid's spectrum. All cases shown in Fig.~\ref{fig: Delta z scan} have been observed in experiments with TMDC excitons coupled to MNPs or PCs, yet lacking a microscopic theory, which we provide from the excitonic perspective.

\subsubsection{Weak Coupling \texorpdfstring{($\delta z=30\,\mathrm{nm}$, $\delta z=17\,\mathrm{nm})$)}{30 nm, 17 nm}}
For MNP-TMDC distances where $\delta z$ is approximately equal to or exceeds the effective extensions of the MNP, the exciton-plasmon coupling is weak.
Thus, for $\delta z\geq 30\,$nm, the line shape is qualitatively preserved compared to the uncoupled case.
However, the interaction induces an additional damping, increasing the linewidth and reducing the maximal absorption. This is because the interaction terms in the denominators of Eqs.~\eqref{eq: Q neq 0 polarization} and \eqref{eq: radiative excitonic transition} are imaginary, thus effectively acting like an additional damping to the phonon damping $\gamma$.
This behavior is analogous to the classical coupled oscillator model~\cite{wu_quantum-dot-induced_2010,torma_strong_2014} and has been experimentally observed between TMDC excitons and MNP plasmons in Refs.~\cite{kleemann_strong-coupling_2017,qin_revealing_2020}.

With further decreasing TMDC-PC distance, the absorption line shape changes. The absorption of the hybrid develops two maxima, while the absorption at the original, non-perturbed, resonance energy shrinks. The splitting between the formed maxima can be identified with an effective Rabi energy $\Omega_\text{eff}$. Its distance dependency is depicted in the inset of Fig.~\ref{fig: Delta z scan}b, illustrating a $(\delta z)^{-2}$ proportionality.
To assign the hybrids to a particular coupling regime, the effective Rabi energy $\Omega_\text{eff}\neq 0$ is compared with the linewidth $\gamma_\text{ex}$ and $\gamma_\text{pl}$ of the individual constituents~\cite{limonov_fano_2017,torma_strong_2014}.

A coupling strength leading to a peak splitting that is small compared to the linewidths as e.g.~in the $\delta z=17\,$nm case is still assigned to weak coupling and has been observed in experiments with TMDCs coupled to a MNP or a PC in Refs.~\cite{abid_temperature-dependent_2017,petric_tuning_2022,lee_fano_2015,vadia_magneto-optical_2023}.

\clearpage

\onecolumngrid

\begin{figure}[b]
 \begin{center}
\includegraphics[width=1.0\linewidth, trim={0 .4cm 0 0}]{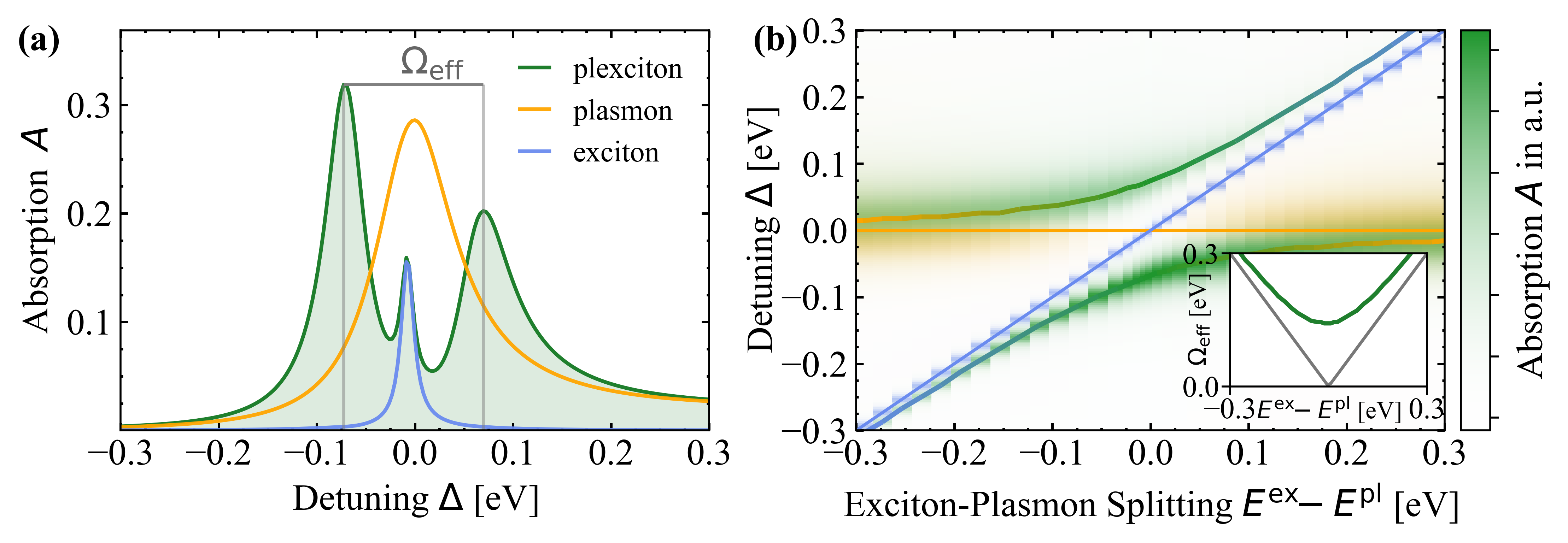}
 \end{center}
 \caption{\textbf{(a) Absorption spectrum} at liquid nitrogen temperature $T=77\,$K of the hybrid structure (green) compared to the uncoupled absorption of the TMDC (blue) and the 2D PC (yellow) for $E^\text{ex}=E^\text{pl}$. The effective Rabi energy $\Omega_\text{eff}\approx 140\,$meV is identified as the energetic separation between the two outer plexcitonic resonances.
 \textbf{(b) Dispersion} of the plexciton branches depending on the excitonic resonance energy $E^\text{ex}$. The upper and lower branches are highlighted with solid lines as a guide to the eye. The hybrid spectrum shows an exciton-plasmon splitting, where plasmon (yellow) and exciton (blue) hybridize to plexcitonic modes (green). The horizontal and main diagonal mark the undisturbed plasmon (yellow) respectively exciton (blue) resonances. \textbf{Inset:} Splitting $\Omega_\text{eff}$ (green) between plexcitonic branches depending on the difference between the resonance energies $E^\text{ex}-E^\text{pl}$. The gray linear plot indicates the splitting without any interaction.\\}
\label{fig: peaksplitting}
\end{figure}
\twocolumngrid

\subsubsection{Strong Coupling \texorpdfstring{($\delta z=11\,{\rm nm}$)}{z = 11 nm}}

Further decreasing the distance of TMDC and PC allows one to enter the strong coupling regime as $\Omega_\text{eff} > \gamma_\text{pl}+\gamma_\text{ex}$ with peak splittings up to $\Omega_\text{eff}\approx 140\,\text{meV}$. 
In this case, mathematically, in Eqs.~(\ref{eq:conicEVProblem}-\ref{eq: Q neq 0 polarization}), a real part is added to the excitonic eigenenergy, thus changing the root of the denominator in Eq.~\eqref{eq: Q neq 0 polarization}, and consequently the plexcitonic resonance energies. Experimentally, strong coupling has been observed up to room temperature \cite{kleemann_strong-coupling_2017,zhang_steering_2021,geisler_single-crystalline_2019,hou_simultaneous_2022,wen_room-temperature_2017,qin_revealing_2020,cuadra_observation_2018,zheng_manipulating_2017,zhang_observation_2023} for TMDC-MNP hybrids with effective Rabi energies similar to the values shown in Fig.~\ref{fig: Delta z scan}.

It would be worth also considering the possibility of ultra-strong coupling in a similar platform. However, this limit cannot be addressed by the developed theory in this work due to the constraint $g_{\mathrm{eff}}/E^{\mathrm{pl}}<0.1$, since the rotating wave approximation is applied (Sec.~\ref{SecTheo}).


\subsection{Low Temperatures}
To study low-temperature effects, Fig.~\ref{fig: peaksplitting}a depicts the absorption spectrum in the strong coupling case for coinciding exciton and plasmon resonance $E^\text{ex}=E^\text{pl}$ at liquid nitrogen temperature $T=77\,$K.
In addition to the typical strong coupling spectrum, {\it we find a third peak at the undisturbed exciton resonance}.
At room temperature (Fig.~\ref{fig: Delta z scan}a) this peak is not well resolved due to the increased damping.
To investigate the dispersion of the three-peak spectrum, we numerically vary the exciton energy around the plasmon resonance in Fig.~\ref{fig: peaksplitting}b. The color coding breaks the full absorption of the hybrid down to exciton (blue), plasmon (yellow) and the hybridized plexciton contribution (green). The individual plasmon and exciton dispersions are indicated by the horizontal line and the main diagonal, respectively.
The avoided-crossing behavior of the upper and lower branch substantiates that the system is indeed in the strong coupling regime. However, the spectral position of the additional middle peak coincides with the unperturbed exciton energy $E^\text{ex}$. The inset depicts the energy separation of the hybrid branches, where the minimum is the effective Rabi energy of exciton and plasmon. For comparison, the linear gray plot indicates the energy separation between exciton and plasmon without any interaction.

\subsubsection{Interpretation of the Bright Excitonic Mode}

The additional peak in Figs.~\ref{fig: peaksplitting}a and \ref{fig: peaksplitting}b stems from the bright excitonic transition $\mathbf{p}_{{\bQp}}$ with in-plane momenta ${\bQp}$ inside the light cone, see Eq.~\eqref{eq: quasi-static condition} and Fig.~\ref{fig: scheme theory}. It is caused by the qualitative difference between the exciton-plasmon coupling strength for bright and momentum-dark excitons that was already discussed in Sec.~\ref{sec: theory explanation}.
With our restriction to PCs that fulfill Eq.~\eqref{eq: dense PC approximation}, the only non-vanishing contribution to the third peak is $\mathbf{p}_{{\bQp}=\mathbf{0}}$, Eq.~\eqref{eq: radiative excitonic transition}, meaning it constitutes a weakly coupled bright excitonic mode.

The appearance of the bright excitonic mode in Fig.~\ref{fig: peaksplitting}a can be traced back to a qualitative difference of our description to previous models~\cite{carlson_strong_2021,denning_quantum_2022,goncalves_plasmon-exciton_2018,wu_quantum-dot-induced_2010}. It results from the geometry of the considered subsystems:
the PC consists of discrete dipoles that feature scattering processes. This allows the plasmon to couple to excitonic transitions outside the light cone via the plasmon-enhanced electric near-field.
In contrast, the TMDC facilitates a continuous translational invariance. Therefore, it is necessary to distinguish between excitonic transitions with in-plane momenta in- and outside the light cone, providing that $\mathbf{p}_{{\bQp}=\mathbf{0}}$ does not couple to the plasmon-enhanced near-field but only to its $\bQp=\mathbf{0}$ Fourier mode.

As plasmons are popular for the amplification of the electric field in their vicinity, the near-field exciton-plasmon interaction is much stronger than the radiative (far-field) contribution.
However, the ${\bQp}=\mathbf{0}$ Fourier component is the propagating mode, also transferring the plasmon response to the far-field. It is physically intuitive, due to the argument of energy conservation, that this mode cannot provide a field enhancement and thus yields a reduced coupling strength to the TMDC excitons.

In Fig.~\ref{fig: peaksplitting}a, we find a subtle damping and broadening due to the radiative exciton-plasmon interaction for the excitonic mode compared to the unperturbed excitonic TMDC absorption (blue), similar to the weak exciton-plasmon coupling regime.

 \subsubsection{Visibility of the Bright Excitonic Mode}
\begin{figure}[htb]
 \begin{center}
\includegraphics[width=\linewidth, trim={0 .6cm 0 0}]{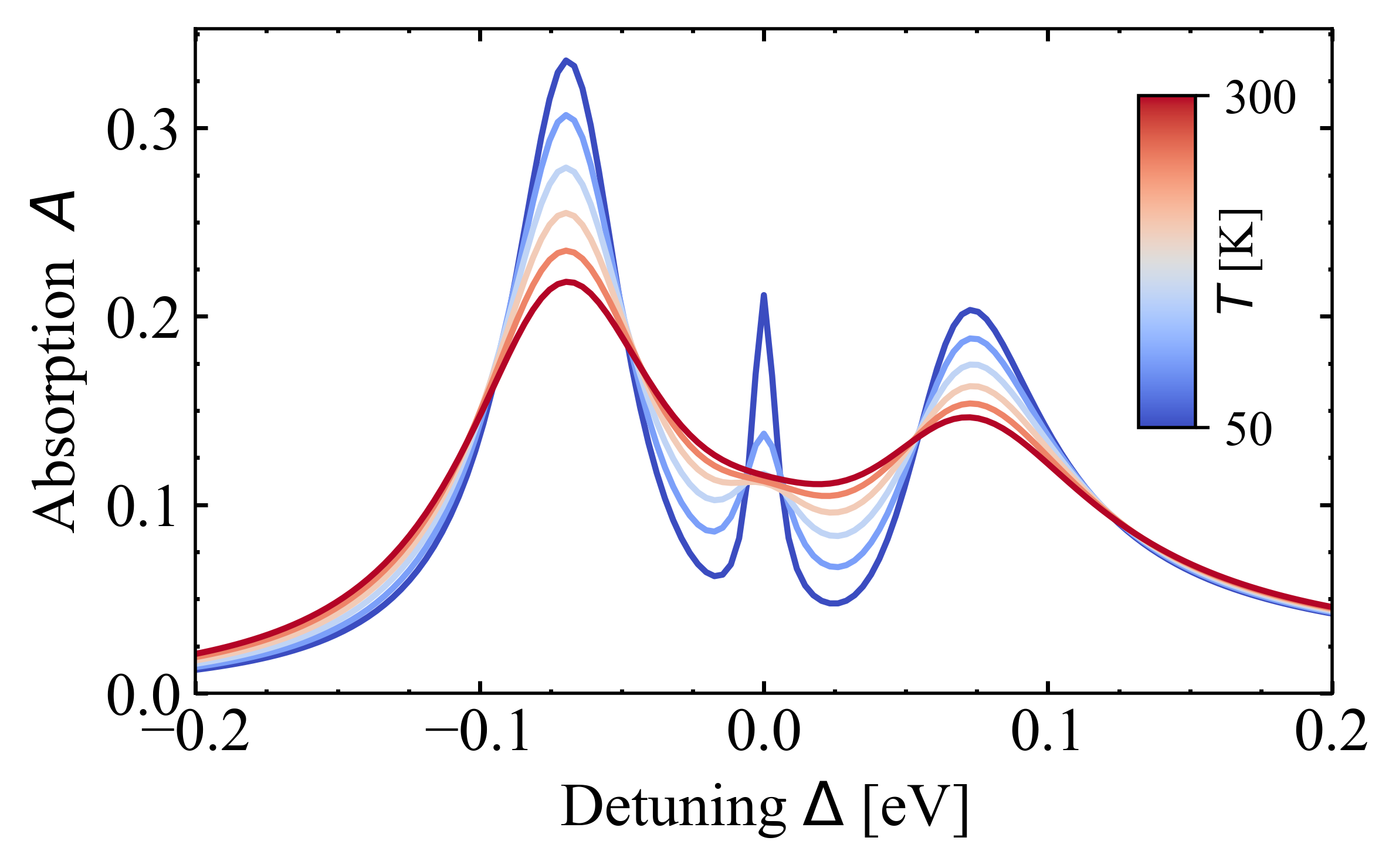}
 \end{center}
 \caption{\textbf{Absorption for different temperatures} $T$ between 50 and 300\,K in the case of coinciding exciton and plasmon resonance. For room temperature $T=300\,$K (red), the hybrid features a typical strong coupling spectrum.}
 \label{fig: T scan}
\end{figure}

Comparing Fig.~\ref{fig: Delta z scan} and Fig.~\ref{fig: peaksplitting}, we find that the visibility of the weakly coupled excitonic mode strongly depends on the temperature. This trend is analyzed in Fig.~\ref{fig: T scan}. It illustrates the hybrid absorption spectrum for several temperatures between $T=50\,$K and $T=300\,$K with the excitonic mode clearly visible for low temperatures. With increasing temperature, the absorption peak shrinks until the increasing exciton damping reaches values of approximately half the peak splitting at room temperature. The larger the peak splitting $\Omega_\text{eff}$, that in turn characterizes the coupling strength, the smaller is the plexcitonic absorption contribution at $\Delta=0$, cf.~Fig.~\ref{fig: Delta z scan}, which improves the visibility of the excitonic mode.

The bright excitonic mode, beyond the usually found peak splitting \cite{abid_resonant_2016,abid_temperature-dependent_2017,lee_fano_2015,petric_tuning_2022, kleemann_strong-coupling_2017, geisler_single-crystalline_2019, wen_room-temperature_2017, zheng_manipulating_2017,goncalves_plasmon-exciton_2018}, was recently observed in Ref.~\cite{vadia_magneto-optical_2023} at cryogenic temperatures. Measured was the differential reflection of a WSe$_2$-monolayer, encapsulated in hBN, on top of a PC with lattice constant $a=300\,$nm consisting of gold nanodisks with radii $r_x=r_y\approx60\,$nm, $r_z\approx8\,$nm and TMDC-PC distance $\delta z = 12\,$nm.
To the best of our knowledge, this was the first observation of the additional excitonic mode,
since most experimental works operate either at room temperature or with the interaction strength too small.

\subsubsection{Comparison to Previous Models}
Previous theoretical analysis was restricted to a phenomenological coupled oscillator model~\cite{wu_quantum-dot-induced_2010} and the Jaynes-Cummings model~\cite{torma_strong_2014} applied on TMDC-MNP and PC hybrids~\cite{petric_tuning_2022,geisler_single-crystalline_2019,abid_resonant_2016,abid_temperature-dependent_2017,cuadra_observation_2018,hou_simultaneous_2022,kleemann_strong-coupling_2017,liu_strong_2016,qin_revealing_2020,vadia_magneto-optical_2023,zheng_manipulating_2017,zhang_observation_2023} or to near-field excitation involving quasinormal modes~\cite{carlson_strong_2021,denning_quantum_2022,denning_cavity-induced_2022}. Numerical approaches using Maxwell solvers have been used to confirm theoretical and experimental results at room temperature \cite{carlson_strong_2021,petric_tuning_2022,kleemann_strong-coupling_2017,geisler_single-crystalline_2019,zheng_manipulating_2017,liu_strong_2016,vadia_magneto-optical_2023,li_tailoring_2018,wen_room-temperature_2017,zhu_electroluminescence_2023,zhang_observation_2023}.
The theory developed in this contribution  provides a solution for the electric near- and far-field explicitly and exceeds the limits of phenomenological models by giving a microscopic description of the TMDC excitons by including the exciton center-of-mass momentum to derive the coupling strength.

When included in the coupled oscillator model, the additional excitonic mode was described as a third oscillator and assigned to excitons spatially separated from the electric field hot spots caused by a single MNP~\cite{geisler_single-crystalline_2019} or a PC~\cite{vadia_magneto-optical_2023}; this phenomenological explanation is consistent with the developed microscopic theory, since the strong coupling is restricted to ${\bQp}\neq \mathbf{0}$ Fourier modes, corresponding to spatially localized states~\cite{salzwedel_spatial_2023}.
However, the ${\bQp}= 0$ mode, responsible for the bright excitonic mode, constitutes a constant spatial distribution in real space.
Our analysis therefore shows that the additional excitonic mode $P^\text{ex}_{{\bQp}= 0}$ is uniformly distributed also in the vicinity of the MNPs.

Another interesting feature of the absorption spectra is the asymmetry of the spectral peaks line shape with respect to width and height. The asymmetry is also covered by phenomenological models \cite{wu_quantum-dot-induced_2010,torma_strong_2014} and is consistently observed in experiments \cite{cuadra_observation_2018,geisler_single-crystalline_2019,kleemann_strong-coupling_2017,liu_strong_2016,petric_tuning_2022,qin_revealing_2020,wen_room-temperature_2017}. The reason for the asymmetry are different line shapes, line widths and dipole moments of exciton and plasmon.
Within our microscopic model, another factor contributes to the observed asymmetry.
The momentum-dark excitons responsible for the peak-splitting exhibit energy distributions up to $50\,$meV on the exciton dispersion, Eq.~\ref{eq: exciton dispersion}, well above the bright exciton state.
Consequently, these excitons, that facilitate the strong exciton-plasmon coupling, are detuned from the joint plasmon and bright exciton resonance.
The magnitude of this energy difference depends on the dielectric environment, cp.~Eq.~\ref{eq: exciton dispersion}, and the distance $\delta z$, as discussed in sec.~\ref{sec: theory explanation}.
The detuning leads to an incomplete hybridization between plasmons and excitons. Consequently, the peak at lower energies effectively contains more plasmonic contributions which results in a stronger absorption.

\section{Conclusions}
We have introduced a self-consistent theory
that highlights the extraordinary optical properties that can be achieved by combining ultrathin semiconductors, such as transition metal dichalcogenides (TMDCs), with plasmonic crystals (PCs) composed of metal nanoparticles. Our approach yields new analytical insights into exciton-plasmon coupling and allows to significantly reduce the numerical costs compared to general Maxwell solvers, such as ANSYS Lumerical (finite-difference time-domain) or COMSOL (finite element).

We developed a theoretical formalism to describe the coupling between a TMDC and a 2D PC, mediated by the self-consistently solved electric field.
More explicitly, we considered the optical response of collective plasmons of 2D PCs within the excitonic Bloch equation. The arising effective intra- and intervalley exciton-exciton interactions for momentum-dark excitons have been decoupled in the eigenbasis of the quasi-static Green's dyadic into excitonic branches with conic respectively parabolic dispersion. Any near-field effects appeared to be incorporated via the conic contribution. In contrast, momentum-bright excitons show a qualitatively different behavior, as the coupling strength is proportional to the negligible center-of-mass momenta.
Nevertheless, signatures of the momentum-dark excitons are visible in the far-field signal due to a second scattering process at the PC.

By evaluating the absorption for different TMDC-MNP separations, we were able to compare different coupling strengths and observe weak and strong coupling line shapes, the latter obeying an avoided-crossing dispersion behavior.
We also find a new feature in the hybrid spectrum at low temperatures beyond the phenomenological model of two coupled oscillators: the momentum-bright TMDC excitons do not participate in the strong coupling due to their reduced coupling strength but emit undisturbed into the far-field.





\begin{acknowledgments}
The authors thank financial support from the Deutsche Forschungsgemeinschaft (DFG) through SFB 951 Project
No. 182087777, Project SE 3098/1-1 (R.S. and M.S.) Project No. 432266622, DPG Sachbeihilfe No.~504656879 (S.R.), the European Commission through the Consolidator Grant DarkSERS (772108)
and the Natural Sciences and Engineering Research Council of Canada.
We also acknowledge support from the 
Alexander von Humboldt Foundation through a Humboldt Research Award (S.H., A.K.).
\end{acknowledgments}

\appendix
\section{Material and Model Parameters\label{app: parameters}}
The parameters for numerical computations are mainly given in semiconductor units  to improve the numerical accuracy which use fs and nm as a measure of time and length, eV for energy and the elementary charge $e$. \\

\begin{table}[h]
	\centering
	\caption{universal constants in semiconductor units}
	\label{table:universal constants}
	\begin{tabular}{c|l}
	$c$ & $299.7925$ nm/fs  \\\hline
	$\hbar$ & $0.658212196$ eV\,fs \\\hline
    $k_B$ & $0.0861745\,$meV/K \\
	\end{tabular}
\end{table}

\begin{table}[h]
	\centering
	\caption{Parameters for MoSe$_2$}
	\label{table: TMDC Material parameters}
	\begin{tabular}{c|l}
	$d$ & $\varphi_{\mathbf{r_\parallel}=\mathbf{0}}\cdot 0.25\,e\,$nm \cite{xiao_coupled_2012} \\\hline
	$\varphi_{\mathbf{r_\parallel}=\mathbf{0}}$ & $0.46\,$nm$^{-1}$	\\\hline
    $M$ &  $6.1\,\text{eV\,fs}^2 \text{nm}^{-2}$ \cite{kormanyos_k_2015}\\\hline
    $c_1$ & $0.091\,$meV/K \cite{selig_excitonic_2016} \\\hline
    $c_2$ & $15.6\,$meV \cite{selig_excitonic_2016}\\\hline
    $\Omega$ & $30\,$meV \cite{selig_excitonic_2016}\\
	\end{tabular}
\end{table}
The excitonic wavefunction $\varphi_{\mathbf{r_\parallel}=\mathbf{0}}$ appears as the solution of the Wannier equation, similar to Refs.~\cite{berghauser_analytical_2014,selig_excitonic_2016} for the chosen dielectric environment.

\begin{table}[h]
	\centering
	\caption{Parameters for the permittivity of gold taken from Ref.~\cite{etchegoin_analytic_2006} (converted to semiconductor units)}
	\label{table: Au Material parameters}
	\begin{tabular}{c|l}
	$\epsilon_\infty$ & $1.53$  \\\hline
	$\omega_\text{p}$ & $12.99\,$fs$^{-1}$  \\\hline
	$\omega_1$ & $4.02\,$fs$^{-1}$	\\\hline
	$\omega_2$ & $ 5.69 \,$fs$^{-1}$	\\\hline
	$\Gamma_1$ & $0.82  \,$fs$^{-1}$	\\\hline
	$\Gamma_2$ & $2.00  \,$fs$^{-1}$	\\\hline
	$A_1$ & $0.94$	\\\hline
	$A_2$ & $1.36$	\\\hline
	$\phi_1$ & $-\pi/4$	\\\hline
	$\phi_2$ & 	$-\pi/4$ \\
	\end{tabular}
\end{table}

\begin{table}[h!]
	\centering
	\caption{Parameters for temperature dependent linewidth of gold taken from Ref.~\cite{liu_reduced_2009} (converted to semiconductor units)}
	\label{table: Au linewidth Material parameters}
	\begin{tabular}{c|l}
    $b$ & $0.6329 \,\text{eV}^{-1}$ \\\hline
    $\gamma_0$ & $0.0219\,$eV \\\hline
    $\Theta$ & $185\,$K\\
	\end{tabular}
\end{table}


\bibliography{Bib}
\end{document}